\documentclass[twocolumn]{aastex63}

\usepackage[T1]{fontenc} 
\usepackage[utf8]{inputenc} 

\shorttitle{The Warm Neptune TOI-2076 b has a Low Obliquity}

\newcommand{\unit}[1]{\ensuremath{\, \mathrm{#1}}}
\newcommand{\respsi}{$\psi=18_{-9}^{+10\:\circ}$}
\newcommand{\respsival}{$18_{-9}^{+10}$} 
\newcommand{\reslambda}{$\lambda=-3_{-15}^{+16\:\circ}$} 
\newcommand{\reslambdaval}{$-3_{-15}^{+16}$}
\newcommand{\resvsini}{$v \sin i = 5.27_{-0.29}^{+0.24} \unit{km/s}$}   
\newcommand{\resvsinival}{$5.27_{-0.22}^{+0.24}$}
\newcommand{\resistarval}{$ 79_{-11}^{+8}$} 
\newcommand{\resistar}{$i_\star = 79_{-11}^{+8\:\circ}$}

\newcommand{\resslopeval}{$89.7 \pm 2.4$}

\newcommand{\PSUAA}{Department of Astronomy \& Astrophysics, 525 Davey Laboratory, The Pennsylvania State University, University Park, PA, 16802, USA}
\newcommand{\PSUCEHW}{Center for Exoplanets and Habitable Worlds, 525 Davey Laboratory, The Pennsylvania State University, University Park, PA, 16802, USA}
\newcommand{\Princeton}{Department of Astrophysical Sciences, Princeton University, 4 Ivy Lane, Princeton, NJ 08540, USA}
\newcommand{\SAGAN}{NASA Sagan Fellow}
\newcommand{\PSETI}{Penn State Extraterrestrial Intelligence Center, 525 Davey Laboratory, The Pennsylvania State University, University Park, PA, 16802, USA}
\newcommand{\UA}{Steward Observatory, The University of Arizona, 933 N.\ Cherry Ave, Tucson, AZ 85721, USA}

\newcommand{\STScI}{Space Telescope Science Institute, 3700 San Martin Dr, Baltimore, MD 21218, USA}
\newcommand{\JHU}{Department of Physics and Astronomy, Johns Hopkins University, 3400 N Charles St, Baltimore, MD 21218, USA}

\newcommand{\JPL}{Jet Propulsion Laboratory, California Institute of Technology, 4800 Oak Grove Drive, Pasadena, California 91109}
\newcommand{\UCI}{Department of Physics \& Astronomy, The University of California, Irvine, Irvine, CA 92697, USA}
\newcommand{\Carleton}{Carleton College, One North College St., Northfield, MN 55057, USA}
\newcommand{\HWS}{Department of Physics, Hobart and William Smith Colleges, 300 Pulteney Street, Geneva, NY 14456, USA}
\newcommand{\GoddardESAL}{Exoplanets and Stellar Astrophysics Laboratory, NASA Goddard Space Flight Center, Greenbelt, MD 20771, USA}

\begin{document}

\title{NEID Reveals that The Young Warm Neptune TOI-2076 b Has a Low Obliquity}

\correspondingauthor{Robert C. Frazier}
\email{rcf5201@psu.edu}

\author[0000-0001-6569-3731]{Robert C. Frazier}
\affil{\PSUAA}
\affil{\PSUCEHW}

\author[0000-0001-7409-5688]{Guðmundur Stefánsson} 
\affil{\Princeton}
\affil{\SAGAN}

\author[0000-0001-9596-7983]{Suvrath Mahadevan}
\affil{\PSUAA}
\affil{\PSUCEHW}
\affil{ETH Zurich, Institute for Particle Physics \& Astrophysics, Switzerland}

\author[0000-0001-7961-3907]{Samuel~W.~Yee}
\affiliation{\Princeton}

\author[0000-0003-4835-0619]{Caleb I. Ca\~nas}
\affil{NASA Goddard Space Flight Center, 8800 Greenbelt Road, Greenbelt, MD 20771, USA}

\author[0000-0002-4265-047X]{Joshua N.\ Winn}
\affil{\Princeton}

\author[0000-0002-4927-9925]{Jacob Luhn}  
\affil{\UCI}

\author[0000-0002-8958-0683]{Fei Dai}
\affiliation{Division of Geological and Planetary Sciences, California Institute of Technology, 1200 E California Blvd, Pasadena, CA, 91125, USA}
\affiliation{Department of Astronomy, California Institute of Technology, Pasadena, CA 91125, USA}

\author[0000-0002-9365-2555]{Lauren Doyle}
\affiliation{Centre for Exoplanets and Habitability, University of Warwick, Coventry, CV4 7AL, UK}
\affiliation{Department of Physics, University of Warwick, Coventry, CV4 7AL, UK}

\author[0000-0001-8934-7315]{Heather Cegla}
\affiliation{Centre for Exoplanets and Habitability, University of Warwick, Coventry, CV4 7AL, UK}
\affiliation{Department of Physics, University of Warwick, Coventry, CV4 7AL, UK}

\author[0000-0001-8401-4300]{Shubham Kanodia}
\affil{Earth and Planets Laboratory, Carnegie Institution for Science, 5241 Broad Branch Road, NW, Washington, DC 20015, USA}

\author[0000-0003-0149-9678]{Paul Robertson}
\affil{\UCI}

\author[0000-0001-9209-1808]{John Wisniewski}
\affil{George Mason University, Department of Physics and Astronomy, 4400 University Drive, MS 3F3, Fairfax, VA 22030, USA}

\author[0000-0003-4384-7220]{Chad F.\ Bender}
\affil{\UA}

\author[0000-0002-3610-6953]{Jiayin Dong} 
\altaffiliation{Flatiron Research Fellow} 
\affiliation{Center for Computational Astrophysics, Flatiron Institute, 162 Fifth Avenue, New York, NY 10010, USA}

\author[0000-0002-5463-9980]{Arvind F.\ Gupta}
\affil{\PSUAA}
\affil{\PSUCEHW}

\author[0000-0003-1312-9391]{Samuel Halverson}
\affil{\JPL}

\author[0000-0002-6629-4182]{Suzanne Hawley}
\affil{Astronomy Department, Box 351580, University of Washington, Seattle, WA 98195}

\author[0000-0003-1263-8637]{Leslie Hebb}
\affil{\HWS}

\author[0000-0002-5034-9476]{Rae Holcomb}
\affil{\UCI}

\author[0000-0001-7458-1176]{Adam Kowalski}
\affil{Department of Astrophysical and Planetary Sciences, University of Colorado Boulder, 2000 Colorado Ave., Boulder, CO 80305, USA}
\affil{National Solar Observatory, University of Colorado Boulder, 3665 Discovery Drive, Boulder, CO 80303, USA}
\affil{Laboratory for Atmospheric and Space Physics, University of Colorado Boulder, 3665 Discovery Drive, Boulder, CO 80303, USA}

\author[0000-0002-2990-7613]{Jessica Libby-Roberts}
\affil{\PSUAA}
\affil{\PSUCEHW}

\author[0000-0002-9082-6337]{Andrea S.J.\ Lin}
\affil{\PSUAA}
\affil{\PSUCEHW}

\author[0000-0003-0241-8956]{Michael W.\ McElwain}
\affil{\GoddardESAL} 

\author[0000-0001-8720-5612]{Joe P.\ Ninan}
\affil{Department of Astronomy and Astrophysics, Tata Institute of Fundamental Research, Homi Bhabha Road, Colaba, Mumbai 400005, India}

\author[0000-0003-0412-9314]{Cristobal Petrovich}
\affiliation{Instituto de Astrofísica, Pontificia Universidad Católica de Chile, Av. Vicuña Mackenna 4860, 782-0436 Macul, Santiago, Chile}
\affiliation{Millennium Institute for Astrophysics, Chile}

\author[0000-0001-8127-5775]{Arpita Roy}
\affil{\STScI}
\affil{\JHU}

\author[0000-0002-4046-987X]{Christian Schwab}
\affil{School of Mathematical and Physical Sciences, Macquarie University, Balaclava Road, North Ryde, NSW 2109, Australia}

\author[0000-0002-4788-8858]{Ryan C. Terrien}
\affil{\Carleton}

\author[0000-0001-6160-5888]{Jason T.\ Wright}
\affil{\PSUAA}
\affil{\PSUCEHW}
\affil{\PSETI}

\begin{abstract}
TOI-2076 b is a sub-Neptune-sized planet ($R=2.39 \pm 0.10  \unit{R_\oplus}$) that transits a young ($204 \pm 50 \unit{MYr}$) bright ($V = 9.2$) K-dwarf hosting a system of three transiting planets. Using spectroscopic observations with the NEID spectrograph on the WIYN 3.5 m Telescope, we model the Rossiter-McLaughlin effect of TOI-2076 b, and derive a sky-projected obliquity of $\lambda=-3_{-15}^{+16\:\circ}$. Using the size of the star ($R=0.775 \pm0.015  \unit{R_\odot}$), and the stellar rotation period ($P_{\mathrm{rot}}=7.27\pm0.23$ days), we estimate an obliquity of $\psi=18_{-9}^{+10\:\circ}$  ($\psi < 34^\circ$ at 95\% confidence), demonstrating that TOI-2076 b is on a well-aligned orbit. Simultaneous diffuser-assisted photometry from the 3.5 m Telescope at Apache Point Observatory rules out flares during the transit. TOI-2076 b joins a small but growing sample of young planets in compact multi-planet systems with well-aligned orbits, and is the fourth planet with an age $\lesssim 300$ Myr in a multi-transiting system with an obliquity measurement. The low obliquity of TOI-2076 b and the presence of transit timing variations in the system suggest the TOI-2076 system likely formed via convergent disk migration in an initially well-aligned disk.
\end{abstract}

\section{Introduction}
\label{sec:intro}
Stellar obliquity---the angle between the stellar rotation axis and the planet orbital axis---is a powerful probe of the dynamical formation histories of planetary systems \citep{albrecht2022}. Stellar obliquities have most successfully been measured with the Rossiter-McLaughlin (RM) Effect \citep{rossiter1924,mclaughlin1924}, which relies on measuring spectral line distortions approximated as radial velocity (RV) shifts during as a companion transits the host star. RM measurements made over the last two decades have revealed a broad distribution of sky-projected obliquities, $\lambda$, from well-aligned to highly misaligned systems \citep[see][and references therein]{albrecht2022}. 

However, the RM effect is primarily sensitive to the sky projection of the obliquity, $\lambda$, rather than the true 3D angle between the stellar rotation axis and the planet orbital axis, $\psi$\footnote{In the special case where the differential rotation is known or can be measured the RM effect can place a constraint on the three-dimensional obliquity \citep[see e.g.,][]{gaudi2007,sasaki2021}.}. When the sky projection of the obliquity $\lambda$ is combined with knowledge of the stellar inclination, $i_\star$, and the orbital inclination of the planet, the 3D obliquity $\psi$ can be estimated. Recently, through constructing a sample of 3D obliquities, \cite{albrecht2021} revealed a possible exoplanet architectural dichotomy, where hot Jupiters are primarily seen to orbit on either well-aligned orbits, or on close to polar orbits. However, the sample from \cite{albrecht2021} is dominated by hot Jupiters---as such planets are the easiest to measure---and it is unclear if this reflects a property intrinsic to how hot Jupiters form, or if this dichotomy is more broadly seen for other types of planetary systems.

With the advent of next-generation RV spectrographs, we are capable of measuring the obliquities of smaller planets, and the growing sample of smaller planets with obliquity measurements hints that the dichotomy might also be seen for such planets \citep{stefansson2022}. However, additional observations are needed to increase the size of the small sample ($\lesssim$10). In this context, observations of low-mass planets ($\lesssim0.3 \unit{M_{J}}$) in young systems ($< 1 \unit{Gyr}$) with precisely determined ages are particularly valuable, as they can help yield insights into the possible timescales involved in the different processes that are invoked to excite obliquities---such as planet-planet scattering \citep{rasio1996}, Von Zeipel-Lidov-Kozai oscillations \citep[e.g.,][]{fabrycky2007,naoz2016}, or secular resonance crossings with a dissapearing disk and a massive outer companion \citep{petrovich2020}---or dampen them through tidal interactions \citep[see discussion in][]{albrecht2012}.

In this letter, we measure the stellar obliquity of TOI-2076 b, a warm sub-Neptune transiting a young \citep[$204 \pm 50 \unit{MYr}$;][]{osborn2022} and active K-dwarf. The TOI-2076 planetary system---which hosts at least three transiting planets---was discovered by \cite{hedges2021} using two sectors of photometric data from the Transiting Exoplanet Survey Satellite \cite[TESS;][]{ricker2015}. The system was further studied by \cite{osborn2022} using the same two sectors in combination with data from the CHEOPS satellite \citep{benz2021}, and revealed Transit Timing Variations (TTVs) at the $\sim$10 min level. The known planets in the system have orbital periods of $P_b = 10.4 \unit{days}$, $P_c = 21.0 \unit{days}$, and $P_d = 35.1 \unit{days}$, and radii of $R_b = 2.5 \unit{R_\oplus}$, $R_c = 3.5 \unit{R_\oplus}$ and $R_d = 3.2 \unit{R_\oplus}$), for planets b, c, and d, respectively \citep{osborn2022}. Further, \cite{zhang2022} recently claim evidence of He 10830\AA\ absorption during the transit of TOI-2076 b, suggesting that the atmosphere is experiencing atmospheric evaporation, though with additional observations \cite{gaidos2022} caution that the absorption signature is most likely due to variability of the Helium line due to stellar activity of the young star.

To constrain the obliquity of TOI-2076 b, we obtained spectroscopic in-transit observations using the NEID spectrograph \citep{schwab2016,halverson2016} on the WIYN 3.5m Telescope at Kitt Peak National Observatory, which demonstrate that TOI-2076 b is on a well-aligned orbit. In addition, diffuser-assisted photometry using the Engineered Diffuser on the ARC 3.5 m Telescope at Apache Point Observatory (APO) reveal no flare events during transit that otherwise could complicate the RM analysis. TOI-2076 b joins a growing sample of warm Neptunes orbiting young stars on well-aligned orbits.

\section{Observations}
\label{sec:obs}

\subsection{TESS}
The Transiting Exoplanet Survey Satellite \citep[TESS;][]{ricker2015} observed TOI-2076 in three sectors: Sector 16 (September 12 to October 6 2019), Sector 23 (March 19 to April 15 2020), and Sector 50 (March 26 to April 22 2022) with the TESS two-minute cadence. TESS uses four CCD cameras to observe in the 600 nm to 1000 nm range, each with a $24 \times 24$ deg field of view, resulting in a combined $24 \times 96$ deg field of view. The TESS data were processed by the Science Processing Operations Center \citep[SPOC;][]{jenkins2016} which removes systematic errors, and extracts the photometry. We retrieved the SPOC photometric data of TOI-2076 using the \texttt{lightkurve} \citep{lightkurve} package. We analyzed the Presearch Data Conditioning Single Aperture Photometry (PDCSAP) lightcurve, which contains systematics-corrected data using the algorithms originally developed for the \textit{Kepler} data analysis pipeline. We removed 12,495 data points with TESS non-zero quality flags (e.g., due to guiding issues, excess stray light), leaving 38,950 data points for subsequent analysis.

\subsection{NEID}
We observed a transit of TOI-2076 b with the NEID spectrograph \citep{schwab2016} on the WIYN 3.5m telescope at Kitt Peak Observatory\footnote{WIYN is a joint facility of the University of Wisconsin–Madison, Indiana University, NSF’s NOIRLab, the Pennsylvania State University, Purdue University, University of California, Irvine, and the University of Missouri.} on the night of 13 Feb 2022 for 5.5 hours. NEID is a high ($R\sim113,000$) resolution spectrograph covering a broad wavelength range from $380-930 \unit{nm}$. In total, using an exposure time of $600 \unit{s}$, we obtained 33 spectra spanning a period of 1 hour before and after the 3.3 hour transit \citep{hedges2021}. The resulting median signal to noise ratio was 105.1 at 5500 \AA. The target rose from airmass 1.90 to 1.02 during the observation.

The NEID spectra were processed with the NEID Data Reduction Pipeline (DRP)\footnote{\url{https://neid.ipac.caltech.edu/docs/NEID-DRP/}}. We extracted the RVs with a custom version of the Spectrum Radial Velocity Analyzer \citep[SERVAL;][]{zechmeister2018} that we have adapted and optimized for NEID data \citep[see][]{stefansson2022}, using all of the available NEID spectra to generate the template for the RV calculation with SERVAL\footnote{We experimented creating a template using derived only from spectra outside of the transit window. Doing so resulted in fully consistent RVs with only a 10cm/s RMS difference between the two different extractions.}. The resulting radial velocities agree well with the RVs from the NEID DRP derived with the cross-correlation functions (CCF) method; the median RV error from the SERVAL pipeline is $0.88 \unit{m/s}$ and it is $1.5 \unit{m/s}$ from the CCF method. We elected to use the SERVAL RVs given their higher RV precision. The RV observations show a feature consistent with the RM effect as well as an upward RV trend that we explore in detail in Section \ref{sec:discussion}.

\subsection{Diffuser-assisted Photometry}
During the spectroscopic transit, we obtained simultaneous photometry using the Engineered Diffuser \citep{stefansson2017} available on the Astrophysical Research Council Telescope Imaging Camera (ARCTIC) instrument \citep{huehnerhoff2016} on the ARC 3.5 m Telescope at Apache Point Observatory. To obtain high precision observations of the bright star, we used the Engineered Diffuser which spreads out the light of the star in a well-defined top-hat shape \citep{stefansson2017} while maintaining a stable PSF throughout the observations. We observed the transit using the SDSS $i^{\prime}$ filter with a short exposure time of 8 seconds due to the brightness of the host star. We used ARCTIC’s $2 \times 2$ binning mode, resulting in a gain of 2.0 e/ADU and a plate scale of $0.22 \unit{\arcsec/pixel}$. The target rose from airmass 1.72 to 1.01 during the observations. Wispy clouds during observation caused transparency fluctuations throughout the observations, which impacted the photometry.

To extract the photometry from the ARCTIC data, we used the AstroImageJ \citep{collins2017} software following the procedures in \cite{stefansson2017}, including bias and flat-field corrections. Prior to final aperture selection and analysis of the photometry, we used the \texttt{astroscrappy} \citep{astroscrappy} code to correct for cosmic rays and other charged particle events. We experimented with using a number of different apertures. For the final light curve analyzed in this work, we used an aperture size of 25 pixels (5.5”) and sampled the background light with an annulus with an inner radius of 45 pixels (9.9”) and an outer radius of 65 pixels (14.3”) around the star as this resulted in the lowest root mean square errors. The six reference stars that resulted in the smallest error in the transit model were all substantially fainter than the target, by about 230 times on average. The observations show a transit-like feature consistent with the expected depth and duration of the transit at the expected time, and reveal no large flares during the transit, which otherwise could complicate the RM analysis.

\section{Stellar Parameters}
\label{sec:stellar}
Table \ref{tab:stellarparam} shows the parameters of the host star TOI-2076 used in this work. We adopt the stellar rotation period from \cite{hedges2021}, $P_{\mathrm{rot}}=7.27 \pm0.23 \unit{days}$, which is precisely determined using long-term ground-based photometry from the Kilodegree Extremely Little Telescope \citep[KELT;][]{pepper2017}. As an additional measurement of the stellar rotation period, we used the \texttt{SpinSpotter} \citep{holcomb2022} code---which uses the autocorrelation function (ACF) to measure stellar rotation periods. Using \texttt{SpinSpotter} on the three available TESS sectors, we obtain a stellar rotation period of $P_{\mathrm{rot}}=7.251 \pm 0.073 \unit{days}$. This value is consistent with the value in \cite{hedges2021}, although the uncertainty from \texttt{SpinSpotter}---which is estimated as the standard deviation of the spacing between ACF vertexes---is underestimated.

To constrain the stellar spectroscopic parameters, we used the \texttt{SpecMatch-Emp}\footnote{\url{https://github.com/samuelyeewl/specmatch-emp}} \citep{yee2017} code---which constrains stellar spectroscopic parameters from comparing a spectrum of a star to a library of as-observed stars with well-constrained spectroscopic parameters---on the highest S/N NEID spectrum of TOI-2076 on segments of 100 \AA between 5000 and 5800 \AA. From the \texttt{SpecMatch-Emp} analysis, we obtained: $T_{\mathrm{eff}} = 5180 \pm 110 \unit{K}$, [Fe/H]=$-0.01\pm0.09$, and $R_* = 0.79 \pm 0.08 \unit{R_{\odot}}$. To constrain the projected rotational velocity of the star, we used the \texttt{SpecMatch-Synth}\footnote{\url{https://github.com/petigura/specmatch-syn}} code \citep{petigura2015}, which compares the spectrum of a star to a library of theoretical spectra \citep{coelho2005} that can be artificially broadened to obtain estimates of $v \sin i$. In doing so, we obtained $v\sin i = 5 \pm 1 \unit{km/s}$ for TOI-2076. The uncertainties from these codes were internally calibrated using a "leave-one-out" procedure with the empirical library of well-characterized stars and observed spectra, and the uncertainties were found to be robust even at S/N as low as 20 per 1D extracted pixel. To obtain a model-dependent constraint on the mass and radius of the star, we performed a spectral energy distribution (SED) fit using the \texttt{SpecMatch-Emp} values and other available magnitudes and priors available from the literature using the \texttt{EXOFASTv2} code \citep{eastman2019} leveraging the Yonsei-Yale stellar isochrone models. The final values are summarized in Table \ref{tab:stellarparam}.

\begin{deluxetable}{llcc}
\tablecaption{Summary of stellar parameters used in this work. \label{tab:stellarparam}}
\tabletypesize{\scriptsize}
\tablehead{\colhead{Parameter}       &  \colhead{Description}                                  & \colhead{Value}                                       & \colhead{Notes}}
\startdata
$d$                                  &  Distance                                               & $41.963{\displaystyle \pm0.028} \unit{pc}$            & (1)        \\
$P_{\mathrm{rot}}$                   &  Stellar Rotation Period                                & $7.27 \pm0.23  \unit{ days}$                          & (2)        \\
\multicolumn{4}{l}{\hspace{-0.3cm}Spectroscopic Parameters from the NEID spectra:}           \\
$T_{\mathrm{eff}}$                   &  Effective Temperature                                  & $5180{\displaystyle \pm110}\unit{K}$                  & (3)        \\
$\mathrm{[Fe/H]}$                    &  Metallicity                                            & $-0.01{\displaystyle \pm0.09 }$                       & (3)        \\
$R_*$                                &  Radius                                                 & $0.79 \pm0.08 \unit{R_{\odot}}$            & (3)        \\
$v\sin i$                            &  Projected Rotational Velocity                          & $5{\pm 1} \unit{km/s}$                                & (3)        \\
\multicolumn{4}{l}{\hspace{-0.3cm} Model-Dependent Stellar SED and Isochrone fit Parameters:}           \\
$M_*$                                &  Mass                                                   & $0.883{\displaystyle \pm0.017} \unit{M_{\odot}}$      & (3)        \\
$R_*$                                &  Radius                                                 & $0.772_{-0.016}^{+0.015} \unit{R_{\odot}}$            & (3)        \\
$\rho_*$ & Stellar Density & $2.720 \pm 0.165 \unit{g/cm^3}$ & (3) \\
$T_{\mathrm{eff}}$                   &  Effective Temperature                                  & $5201_{-61}^{+66}  \unit{K}$                          & (3)        \\
Age                                  &  Age                                                    & $0.338_{-0.081}^{+0.077} \unit{Gyr}$                  & (3)        \\
$\mathrm{[Fe/H]}$                    &  Metallicity                                            & $0.017_{-0.056}^{+0.077}$                             & (3)        \\
$\log g$                             &  Surface Gravity in cgs units                           & $4.608{\displaystyle \pm0.018}$                       & (3)        \\
\enddata
\tablenotetext{}{References are: (1) Gaia \citep{gaia2018}, (2) \cite{hedges2021}, (3) This work.}
\end{deluxetable}

\begin{deluxetable*}{lccc}
\tablecaption{Summary of priors and resulting posteriors for the photometric analysis. $\mathcal{N}(m,\sigma)$ denotes a normal prior with mean $m$, and standard deviation $\sigma$; $\mathcal{U}(a,b)$ denotes a uniform prior with a start value $a$ and end value $b$; and $J(a,b)$ denotes a log-uniform distribution between a lower limit $a$ and an upper limit $b$. \label{tab:planetparams}}
\tablehead{\colhead{Parameter}&                                \colhead{Description}  &      \colhead{Priors}                  &      \colhead{Posteriors}           }
\startdata
\multicolumn{4}{l}{\hspace{-0.3cm} Juliet Input Parameters:}           \\
              $T_{\mathrm{TESS}_0}$ $(\mathrm{BJD_{TDB}})$ &  Transit Midpoint, $1^{\mathrm{st}}$ transit &      { $\mathcal{N}(2458743.7248,0.042)$}          & $2458743.7183_{-0.0054}^{+0.0045}$  \\
                $T_{\mathrm{TESS}_1}$ & Transit Midpoint,                                    $2^{\mathrm{nd}}$ transit &      { $\mathcal{N}(2458754.080049,0.042)$}          & $2458754.0768_{-0.0026}^{+0.0022}$  \\
                $T_{\mathrm{TESS}_{19}}$ & Transit Midpoint,                                 $20^{\mathrm{th}}$    transit &      { $\mathcal{N}(2458940.474531,0.042)$}          & $2458940.4810_{-0.0020}^{+0.0097}$  \\
              $T_{\mathrm{TESS}_{20}}$ & Transit Midpoint,                                    $21^{\mathrm{st}}$ transit &      { $\mathcal{N}(2458950.82978,0.042)$}          & $2458950.8343_{-0.0020}^{+0.0097}$  \\
                $T_{\mathrm{TESS}_{89}}$ & Transit Midpoint,                                    $90^{\mathrm{th}}$ transit &      { $\mathcal{N}(2459665.341961,0.042)$}          & $2459665.3542_{-0.0052}^{+0.0038}$  \\
               $T_{\mathrm{TESS}_{90}}$ & Transit Midpoint,                                    $91^{\mathrm{st}}$ transit &      { $\mathcal{N}(2459675.69721,0.042)$}          & $2459675.6942_{-0.0018}^{+0.0019}$  \\
                  $T_{\mathrm{TESS}_{91}}$ & Transit Midpoint,                       $92^{\mathrm{nd}}$ transit &      { $\mathcal{N}(2459686.052459,0.042)$}          & $2459686.0510_{-0.0013}^{+0.0014}$  \\
                     $T_{\mathrm{APO}_{85}}$ & Transit Midpoint,                      $86^{\mathrm{th}}$ transit &      { $\mathcal{N}(2459623.920965,0.042)$}          & $2459623.9183_{-0.0016}^{+0.0011}$  \\
                      $R_p/R_*$ &                                       Radius ratio  &      {$\mathcal{U}(0.0,0.1)$}                           & $0.0284_{-0.0010}^{+0.0011}$       \\
                            $e$ &                                       Eccentricity  &      $0.0$                          & $0.0$                               \\
                       $\omega$ &                  Argument of periastron ($^\circ$)  &      $90.0$                       & $90.0$                             \\
                            $b$ &                                   Impact parameter  &      {$\mathcal{U}(0.0,1.0)$}                           &  $0.123_{-0.073}^{+0.077}$             \\
                         $\rho$ &                               Stellar density (cgs) &      {$\mathcal{N}(2.720, 0.165)$ }                     &       $2.712_{-0.110}^{+0.099}$         \\
                          $q_{1\mathrm{TESS}}$ &                     Linear limb darkening parameter &      {   $\mathcal{U}(0.0,1.0)$ }                       &  $0.724_{-0.230}^{+0.186}$            \\
                          $q_{2\mathrm{TESS}}$ &                  Quadratic limb darkening parameter &      {$\mathcal{U}(0.0,1.0)$}                           &  $0.463_{-0.192}^{+0.227}$          \\
                          $M_\mathrm{Dilution,TESS}$ &                  Dilution Factor &      {$1.0$}                           &  $1.0$          \\
                          $M_\mathrm{Flux,TESS}$ &                  Offset Relative Flux &      {$\mathcal{N}(0.0,0.1)$}                           &  $-0.001_{-0.001}^{+0.001}$          \\
                          $\sigma_{W\mathrm{TESS}}$ &                  Jitter (ppm) &      {$\mathcal{J}(1.0,5000.0)$}                           &  $140.7_{-5.1}^{+5.1}$          \\
                          $q_{1\mathrm{APO}}$ &                     Linear limb darkening parameter &      {   $\mathcal{U}(0.0,1.0)$ }                       &  $0.196_{-0.119}^{+0.170}$            \\
                          $q_{2\mathrm{APO}}$ &                  Quadratic limb darkening parameter &      {$\mathcal{U}(0.0,1.0)$}                           &  $0.602_{-0.322}^{+0.256}$          \\
                          $M_\mathrm{Dilution,APO}$ &                  Dilution Factor &      {$1.0$}                           &  $1.0$          \\
                          $M_\mathrm{Flux,APO}$ &                  Offset Relative Flux &      {$\mathcal{U}(0.0,0.1)$}                           &  $0.00246_{-0.00024}^{+0.00026}$          \\
                          $\sigma_{W\mathrm{APO}}$ &                  Jitter (ppm) &      {$\mathcal{J}(1.0,5000.0)$}                           &  $999.30_{-1.02}^{+0.51}$          \\
\multicolumn{4}{l}{\hspace{-0.3cm} Detrending Parameters:}           \\
                          $B_\mathrm{TESS}$ &                  GP Amplitude &      {$\mathcal{J}(10^{-6},1.0)$}                           &  $0.000019_{-0.000005}^{+0.000008}$          \\
                          $C_\mathrm{TESS}$ &                  GP Additive Factor &      {$\mathcal{J}(0.001,1000.0)$}                           &  $0.014_{-0.011}^{+0.058}$          \\
                          $L_\mathrm{TESS}$ &                  GP Length Scale (days) &      {$\mathcal{J}(0.0,100000.0)$}                           &  $11.99_{-3.10}^{+5.19}$          \\
                          $P_\mathrm{TESS}$ &                  GP Period (days) &      {$\mathcal{J}(2.0,10.0)$}                           &  $4.01_{-0.16}^{+0.17 a}$          \\                         
                           $\theta_{0\mathrm{APO}}$ &                  Linear Regressor Coefficient (APO airmass) &      {$\mathcal{U}(-100.0,100.0)$}                           &  $0.002_{-0.002}^{+0.002}$          \\
 \multicolumn{4}{l}{\hspace{-0.3cm} Derived Parameters:}  \\
            $T_{C}$ $(\mathrm{BJD_{TDB}})$ &                                    Transit midpoint &     -         & $2458743.7247_{-0.0022}^{+0.0027}$  \\
            $P$ &                              Orbital period (days)  &      -                      & $10.35527_{-0.00004}^{+0.00003}$  \\
            $a/R_*$ &                             Scaled semi-major axis  &      -                           &  $24.87_{-0.33}^{+0.30}$         \\
            $i$ &                             Inclination ($^\circ$)  &   -    &  $89.72_{-0.18}^{+0.17}$   \\
            $S_{\mathrm{flux}}$ &                                Insolation flux ($S_\oplus$)      &  -   &   $49.4_{-2.7}^{+2.9}$  \\
            $T_{14}$ &                       Full transit duration (days)  &   -    &  $0.1352_{-0.0018}^{+0.0016}$  \\ 
            $ T_{23} $ &                   Interior transit duration (days)  &   -    &  $0.1275_{-0.0017}^{+0.0016}$  \\ 
            $ \tau $ &                                Ingress time (days)  &   -   &  $0.0038_{-0.00016}^{+0.00017}$  \\
            $R_p$ &       Planetary radius $(R_\oplus)$ &    -   &  $2.39_{-0.10}^{+0.10}$    \\   
            $T_{\mathrm{eq},a=0}$ & Planet equilibrium temperature, albedo $a=0$ (K)$^b$  &    -   &  $738.0 \pm 10.0$       \\
\enddata
\tablenotetext{}{$^a$ This value is not equal to the stellar rotation period value which is $7.27 \pm .23$ days as reported in Table \ref{tab:stellarparam}. See discussion in Section \ref{sec:photometric}.}
\tablenotetext{}{$^b$ Calculated assuming the whole surface is the emitting area.}
\end{deluxetable*}

\section{Photometric Analysis}
\label{sec:photometric}
\begin{figure*}[t!]
\begin{center}
\includegraphics[width=1.0\textwidth]{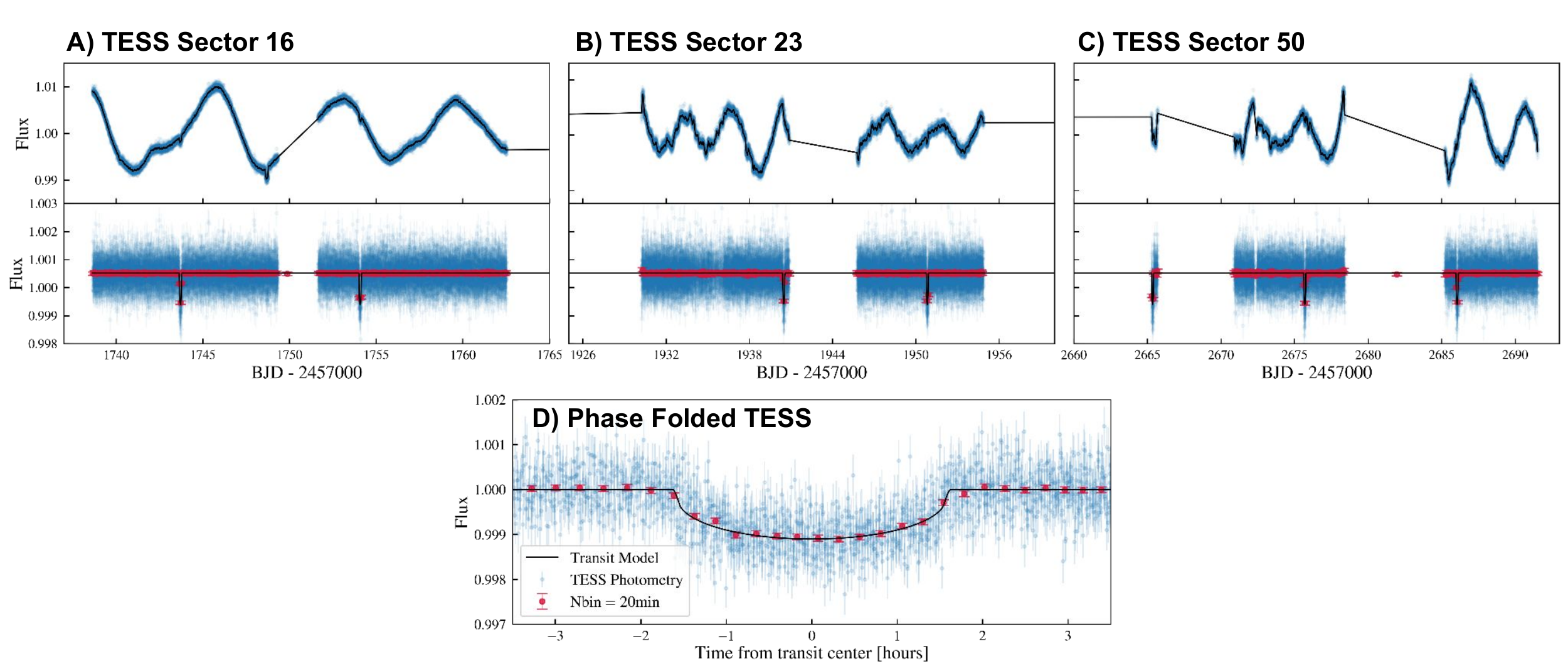}
\vspace{-0.8cm}
\end{center}
\caption{TESS lightcurves of TOI-2076 from A) Sector 16, B) Sector 23, and C) Sector 50. The top row shows the TESS photometry (blue points) with along with a transit model for TOI-2076 b + GP model (red) to account for stellar activity. The middle panel shows the photometry after subtracting the quasi-periodic GP model, revealing the TOI-2076 b transits. Panel D) shows the phase-folded TESS photometry phased to the orbital period of TOI-2076 b, with the transit model overlaid.}
\label{fig:TESS}
\end{figure*}

To precisely constrain the orbital ephemerides of TOI-2076 b---important for the RM effect modeling---we utilized the \texttt{juliet} code \citep{juliet} to perform a fit of the available TESS and diffuser-assisted APO transit photometry. As transit timing variations have been reported in the system with $\sim$10-15min amplitudes \citep{osborn2022,zhang2022}, we leveraged the functionality within \texttt{juliet} to account for TTVs by fitting the individual transit midpoints of the TESS transits and the APO transits separately. Our derived transit midpoints for TOI-2076 b are within $1 \sigma$ with those reported in \cite{osborn2022} and \cite{zhang2022}. For the fit, we used the \texttt{dynesty} dynamic nested sampler \citep{speagle2019} to sample the posteriors and the \texttt{batman} package \citep{kreidberg2015} to generate the light curve models.  The priors and posteriors from the fit are summarized in Table \ref{tab:planetparams}. Figure \ref{fig:TESS} shows the TESS photometry along with the best-fit model and the phase-folded TESS photometry after accounting for the TTVs. The APO photometry and the best-fit model is shown in Figure \ref{fig:APO_RM}.

To remove clear correlated noise signatures seen in the TESS data, we used the 'quasi-periodic` Gaussian Process kernel from the \texttt{celerite} package \citep{Foreman-Mackey2017} available in \texttt{juliet}. Broad uninformative priors were placed on the GP hyper parameters. We note that the GP value $P_{TESS} = 4.01^{+0.17}_{-0.16}$ in Table \ref{tab:planetparams} is different than our adopted value for stellar rotation period of $7.27 \pm 0.23 \unit{days}$ as listed in Table \ref{tab:stellarparam}. We attribute this difference being due to a combination of data gaps, and different spot evolution/behavior seen in the different TESS sectors. The latter two TESS sectors show faster time-scale variability than the first TESS sector, which we attribute to the young and active star likely having developed more spot complexes at different latitudes/longitudes. As noted in Section \ref{tab:stellarparam}, we adopt the $7.27 \pm 0.23 \unit{days}$ value as the rotation period, as that value is derived from 8 years of ground-based monitoring as discussed in \cite{hedges2021}. 

For the ground-based APO data, we observed a smooth trend during the observations. To remove the trend, we employed the linear detrending models available in \texttt{juliet}. We experimented detrending with a number of linear detrending parameters, including the airmass, X, and Y centroid coordinates, time etc. We found that a linear detrending using the airmass parameter yielded the highest quality fit measured from the residual scatter in the photometry after subtracting the detrended transit model from the data. 

For the fit, we placed an informative Gaussian prior on the stellar density of $2.720 \pm 0.165 \unit{g/cm^3}$ (see Table \ref{tab:stellarparam}). We used the ($q_1$, $q_2$) limb-darkening parameterization as described in \cite{kipping2013}. We experimented with a circular and an eccentric fit to model the photometry. Given that we only see a minimal statistical preference of $\Delta \ln (Z)= 0.42$ in favor of the eccentric model and that the posteriors between the two runs were within 1$\sigma$ of each other (the eccentric model yielded a coarse constraint on $e = 0.28^{+0.37}_{-0.23}$ with $e<0.77$ at 95\% confidence), we elected to model the photometry assuming a circular orbit. This agrees with the approach of \cite{hedges2021}. Given the large $a/R_\star \sim 25$ for TOI-2076b, we acknowledge that it is a possibility that additional precise photometric observations could constrain the eccentricity further.

\section{Rossiter McLaughlin Effect}
\label{sec:RM}
To model the RM effect, we used the \texttt{rmfit} package \citep[see][]{stefansson2022}, which implements the RM effect model from \cite{hirano2011}. For the model, we placed informative priors on the transit parameters ($P$, $T_C$, $R_p/R_\star$, $i$, $a/R_\star$) as constrained by the photometric analysis in Section \ref{sec:photometric}. We placed informative priors on the limb-darkening parameters $u_1$ and $u_2$. To arrive at a self-consistent constraint on the 3D obliquity $\psi$, in the RM model, we follow the RM parameterization discussed in \cite{stefansson2022}, which parameterises the RM model in terms of the sky-projected obliquity $\lambda$, stellar inclination ($\cos i_\star$), stellar rotation period ($P_{rot}$), and stellar radius ($R_\star$) as variables in the MCMC sampling. The $v\sin i_\star$ is then estimated as $v \sin i_\star = v \sqrt{1-\cos^2{i_\star}}$, where we estimate the equatorial velocity as $v_{\mathrm{eq}} = v = 2\pi R_\star / P_{\mathrm{rot}}$\footnote{This broadly follows the methodology in \cite{masuda2020} to account for the fact that $v \sin i_\star$ and $v_{\mathrm{eq}}$ are not independent variables. As we are assuming solid body rotation, the equatorial velocity equals the rotational velocity of the star.}. We then estimate the obliquity using,
\begin{equation}
\cos(\psi) = \sin(i_\star)\cos(\lambda)\sin(i) + \cos(i_\star)\cos(i),
\end{equation}
where $i$ is the orbital inclination of the planet. For the final MCMC fit, we ran 100 walkers using the \texttt{emcee} package \citep{dfm2013} for 30,000 steps after removing 2,000 steps as burn-in steps. To determine that the chains were converged, we leveraged a few methods. First, we verified that the Gelman-Rubin (GR) statistic was within $\ll$1\% of unity. However, the GR statistic has limitations to assess convergence, especially when the walkers are not independent \citep[see e.g., discussion in][]{hogg2018}, so in addition to the GR statistic, we follow the suggestion in \cite{hogg2018} and estimated the autocorrelation length of our chains, where the mean autocorrelation length was $\tau_{\mathrm{mean}} = 202$, and the maximum autocorrelation length was $\tau_{\mathrm{max}}=261$. From running 30,000 steps, this ensures that each chain has at least 100 independent samples, which is more than the chain length of $50\tau$ as suggested in the \texttt{emcee} documentation\footnote{See notes here: \url{https://emcee.readthedocs.io/en/latest/tutorials/autocorr/\#autocorr}}. From these lines of evidence combined with visual inspection of the chains suggesting convergence, we conclude the chains are well mixed.

To account for the RV trend seen during the observations, we simultaneously fit the RM effect with a RV slope. We also experimented with adding a quadratic curvature to the slope. However, doing so did not significantly improve the resulting fit where the difference in the Bayesian Information Criterion was $\Delta\mathrm{BIC}=3.0$ in favor of the quadratic model, suggesting only a modest statistical preference for the more complicated model, where both models yielded the same constraints on the key parameters of interest of $v \sin i_\star$ and $\lambda$. Given the low statistical preference, we adopted the simpler linear model. As the semi-amplitude of the planet $K$---which is currently unconstrained as TOI-2076 b does not have a measured mass---is degenerate with the linear RV slope during the short observing baseline, we elected to fix the semi-amplitude of the planet to zero and let the slope parameter fully account for the long-term RV trend during the observations. Table \ref{tab:rmparams} summarizes the input priors and the best-fit posterior values.

\begin{figure*}[t!]
\begin{center}
\includegraphics[width=0.95\textwidth]{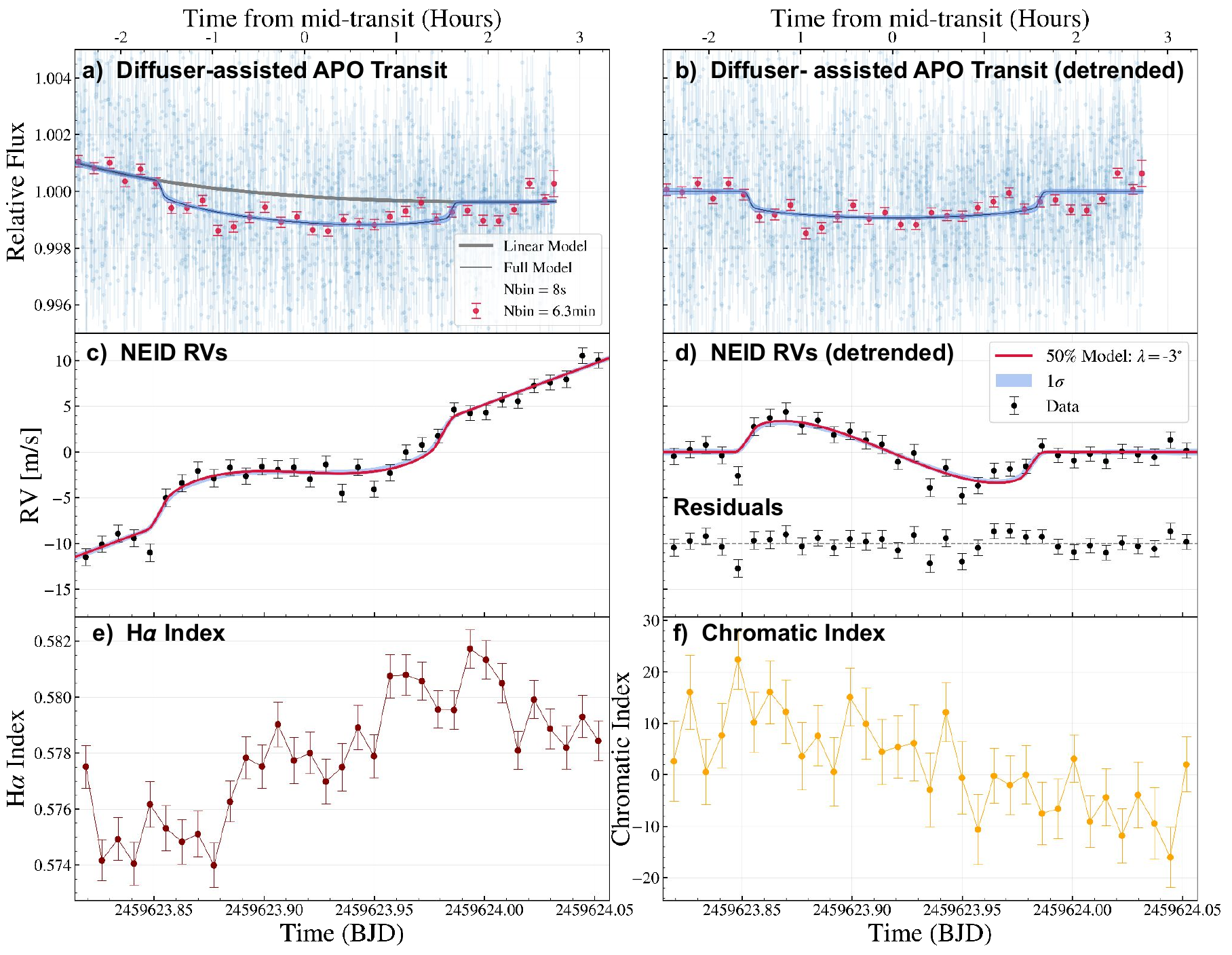}
\vspace{-0.4cm}
\end{center}
\caption{a) Diffuser-assisted photometry from APO with the transit model from the joint APO and TESS fit overlaid in blue. The linear model of the airmass is shown in grey. b) Diffuser-assisted photometry and the joint transit model detrended from the linear model of the airmass. c) NEID RV data during the transit of TOI-2076 b revealing a clear detection of the RM effect. We attribute the positive slope to stellar activity. d) NEID RV data detrended from the overall upward slope and the corresponding residuals. e) H$\alpha$ index from NEID RV observations. f) Chromatic index (CRX) from NEID RV observations. The data are available as data-behind the figure.}
\label{fig:APO_RM}
\end{figure*}

Figure \ref{fig:APO_RM} shows a plot of the data along with the best-fit RM effect model. Our best fit model suggests a sky-projected velocity of \resvsini, a sky-projected obliquity of \reslambda, and a obliquity of \respsi\ ($\psi < 34^{\:\circ}$ at 95\% confidence). The $v\sin i_\star$ value agrees with the $v \sin i_\star= 5\pm 1 \unit{km/s}$ from the spectral-broadening measurements in Table \ref{tab:stellarparam}. The obliquity value of \respsi\ suggests that TOI-2076 b is on a well-aligned orbit.

As an additional test, we fit the RM effect using the more conventional parameterization of $\lambda$ and $v \sin i$ placing uniform priors on $v \sin i_\star$ from 0 to 10km/s, instead of the $\cos i_\star$, $R_\star$, and $P_{\mathrm{rot}}$ parameterization discussed above. In doing so, we obtain a $\lambda$ of $-3_{-14}^{+13\:\circ}$, and $v \sin i_\star = 5.28_{-0.28}^{+0.23} \unit{km/s}$, which agrees with the values above. We elect to use the former parameterization to arrive at a self-consistent constraint of the obliquity, $\psi$.

\begin{deluxetable*}{llccc}
\tablecaption{Summary of priors and resulting posteriors for the RM analysis. $\mathcal{N}(m,\sigma)$ denotes a normal prior with mean $m$, and standard deviation $\sigma$; $\mathcal{U}(a,b)$ denotes a uniform prior with a start value $a$ and end value $b$. \label{tab:rmparams}}
\tablehead{\colhead{~~~Parameter}&                             \colhead{Description}  &      \colhead{Prior} & \colhead{Posterior}}
\startdata
\multicolumn{4}{l}{\hspace{-0.3cm} MCMC Input Parameters:}           \\
 $T_{C}$ $(\mathrm{BJD_{TDB}})$                          &  Transit midpoint                     & $\mathcal{N}(2458743.7247, 0.0025)$ & $2458743.7219 \pm 0.0019$        \\
 $P$                                                     &  Orbital period (days)                & $\mathcal{N}(10.35527,0.00003)$    & $10.355235 \pm 0.00002$         \\
 $R_p/R_*$                                               &  Radius ratio                         & $\mathcal{N}(0.0284, 0.0011)$       & $0.02881 \pm 0.00095 $ \\
 $a/R_*$                                                 &  Scaled semi-major axis               & $\mathcal{N}(24.87, 0.31)$          & $25.07 \pm 0.28 $               \\
 $i$                                                     &  Transit inclination ($^\circ$)       & $\mathcal{N}(89.72, 0.18)$          & $89.65 \pm 0.17$                \\
 $e$                                                     &  Eccentricity                         & 0.                                  & 0                               \\
 $\omega$                                                &  Argument of periastron ($^\circ$)    & 90.                                 & 90.                             \\
 $K$                                                     &  RV semi-amplitude (m/s)              & 0                                   & 0                               \\
 $\gamma$                                                &  NEID RV offset (m/s)                 & $\mathcal{U}(-50,50)$           & $-0.59 \pm 0.18  $              \\
 $u_1$                                                   &  Linear limb darkening parameter      & $\mathcal{N}(0.35,0.1)$             & $0.416_ \pm 0.098$       \\
 $u_2$                                                   &  Quadratic limb darkening parameter   & $\mathcal{N}(0.32,0.1)$              & $0.332 \pm 0.050$               \\
 $\beta$                                                 &  Intrinsic stellar line width (km/s)  & $\mathcal{N}(6.0,1.0)$              & $5.4 \pm 1.0$                   \\
 $\lambda$                                               &  Sky-projected obliquity (deg)        & $\mathcal{U}(-180,180)$             & \reslambdaval                   \\
 $R_\star$                                               &  Radius of star  (\unit{R_{\odot}})                     & $\mathcal{N}(0.772,0.015)$            & $0.774 \pm 0.015$               \\
 $P_{rot}$                                               &  Stellar rotation period (days)       & $\mathcal{N}(7.27,0.23)$           & $7.21 \pm 0.22$                  \\
 $\cos i_\star$                                                &  Cosine of stellar inclination        & $\mathcal{U}(0,1)$                  & $0.2_{-0.14}^{+0.18}$          \\
 $\dot{\gamma} $                                         &  Slope of Radial Velocities (m/s/day) & $\mathcal{U}(-500,500) $            & \resslopeval                    \\
 \multicolumn{4}{l}{\hspace{-0.3cm} Derived Parameters:} \\
 $v \sin i_\star$                                              &  Projected rotational velocity (km/s) & -                                   & \resvsinival                    \\
 $i_\star$                                               &  Stellar inclination (deg)            & -                                   & \resistarval                    \\
 $\psi$                                                  &  Obliquity (deg)                 & -                                   & \respsival                      \\
\enddata
\end{deluxetable*}

\section{Discussion}
\label{sec:discussion}

\subsection{Impact of Stellar Activity on the RM analysis}
TOI-2076 shows clears signatures of stellar activity due to its young age of ($204 \pm 50 \unit{MYr}$). The TESS data shows photometric variations with an amplitude of $\sim$1\% (Figure \ref{fig:TESS}). Additionally, Figure \ref{fig:APO_RM} shows a clear trend in the RVs during the RM observations with an amplitude of $\sim$20 m/s across the full 5.5 hour observing baseline. This trend is in the opposite direction to what we would expect due to the planet-induced stellar radial velocity, which we estimated to have a RV semi-amplitude of $K=2.0_{-0.8}^{+1.5}\unit{m/s}$ (assuming a predicted mass of $6.3_{-2.6}^{+4.5} M_\oplus$ using the mass-radius relation from the \texttt{Forecaster} package \citep{chen2017}). Such inverse trends have been previously reported in RM observations of other young systems: \cite{wirth2021} observed an RV trend of $\sim$75 m/s over a similar baseline ($\sim 19 \unit{m/s/hr}$) in RM observations of the 60 Myr old TOI-942b, and \cite{benatti2019} observed an even larger RV trend of $\sim$150 m/s ($\sim 25 \unit{m/s/hr}$) for the 45 Myr old DS Tuc A b.

\begin{figure*}[t!]
\begin{center}
\includegraphics[width=0.9\textwidth]{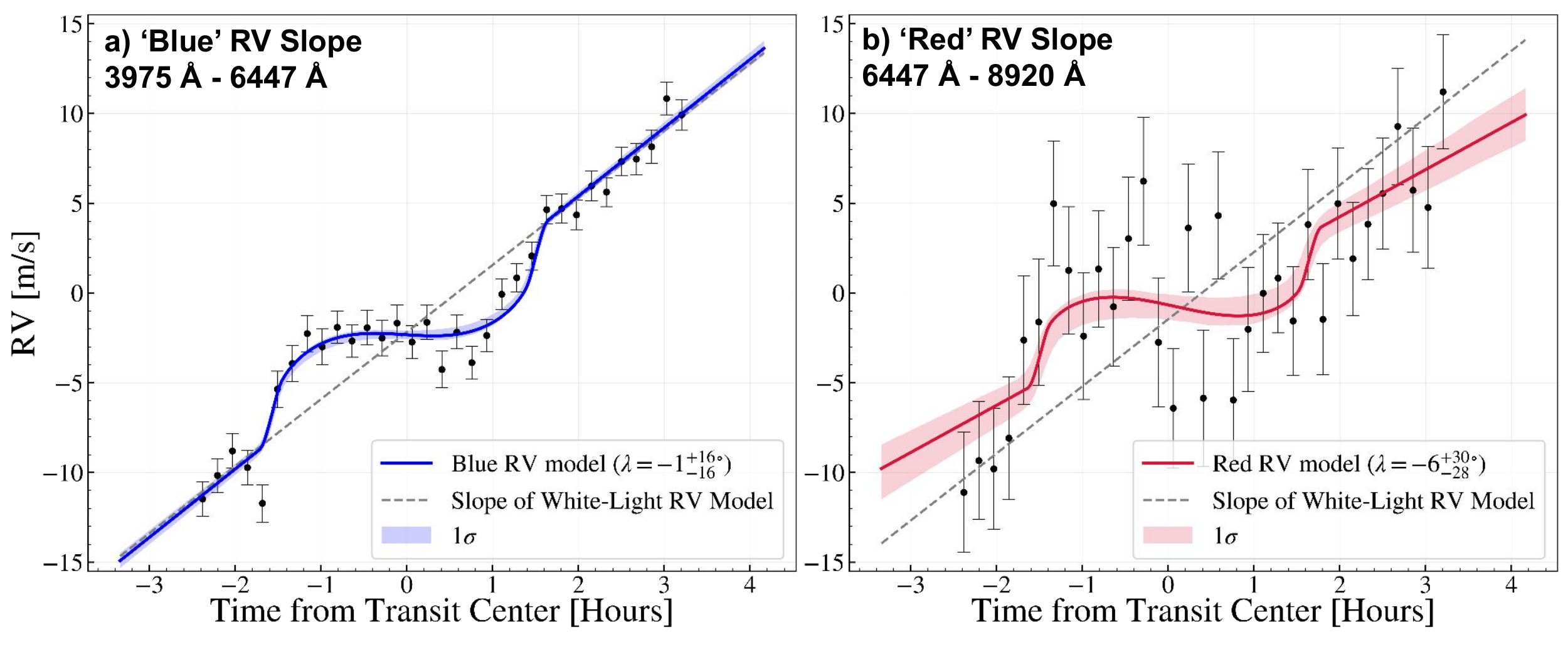}
\vspace{-0.4cm}
\end{center}
\caption{Comparing an RM fit to `blue-only' RVs (a) and `red-only' RVs (b). The RV curve for the `blue-only' RV orders spans orders from 3975--6447 \AA\, and the `red-only' orders span orders from 6447--8920 \AA. For comparison, the slope of the `white-light' RV model used in the full analysis is overplotted as the grey-dashed line in both panels. The slope in the blue orders is steeper than in the red, suggesting the RV slope is due to stellar activity such as a spot on the surface of the star.}
\label{fig:redblue}
\end{figure*}

In addition to the photometry and the radial velocities, Figure \ref{fig:APO_RM} also shows the H$\alpha$ index along with the chromatic index (CRX), both of which show slow trends during the RM observations. We calculated both indices following the definition in \cite{zechmeister2018}. The H$\alpha$ index is particularly sensitive to flares \citep[e.g.,][]{ichimoto1984}, and the lack of flare-like features and/or high-frequency variations suggests that no large flares occurred during the observations, consistent with the photometric observations. The chromatic index, CRX, is defined in \cite{zechmeister2018} as the best-fit straight line slope fitted to order-by-order RVs as a function of wavelength, it is measured in velocity per unit wavelength ratio. A non-zero CRX value signifies that a trend is seen in the order-by-order RVs as a function of wavelength, a strong indication of stellar activity. From the CRX values shown in Figure \ref{fig:APO_RM}f, we see indications of a chromaticity in the RVs. To further examine this behavior, we split the radial velocity data in two groups of `blue' orders (3975--6447\AA) and `red' orders (6447--8920\AA) as seen in Figure \ref{fig:redblue}. Separate RV fits of these `blue' and `red' extractions resulted in RV slopes of $\dot{\gamma}_{\mathrm{blue}} = 91.3_{-2.5}^{+2.4} \unit{m/s/day}$, and $\dot{\gamma}_{\mathrm{red}} = 62.9 \pm 8.5 \unit{m/s/day}$. The lower slope value seen in the red wavelengths is expected if the activity signature is due to a starspot with a differing contrast as a function of wavelength compared to the stellar photosphere, further confirming that the slope is due to activity. Best-fit RM models to the `blue' and `red' RV extractions returned $v\sin i$, $\lambda$ and $\psi$ values consistent with the values (`white-light') reported in Table \ref{tab:rmparams}, suggesting that our treatment of a simple line to remove the RV trend is sufficient to remove the activity signature and does not impact the determination of the obliquity.

To further estimate the expected RV impact of a possible starspot moving on the stellar surface during the RM observations, we used the \texttt{SOAP2.0} code \citep{dumusque2014}. For the starspot simulation, we assumed a stellar inclination of $i_\star = 80^\circ$, consistent with the median value from our RM analysis. We experimented placing different spots of different sizes and found that placing two circular spots with a temperature contrast of $\Delta500 \unit{K}$ of radius 0.23$R_\star$ (covering 2.6\% of the visible hemisphere) at latitudes of 30 degrees, resulted in $\sim1\%$ photometric variations peak-to-valley, consistent with the amplitude of variations seen in the TESS photometry. The expected peak-to-valley RV variations from such spots was $\sim$150 m/s, which would cause a $\sim$19 m/s RV variation during the 5.5 h observing baseline. This is in good agreement with the RV trend we see in Figure \ref{fig:APO_RM}c. We note that this is not a unique solution, as due to degeneracies between different spot parameters, including spot size, latitude, and contrast, there is a good possibility that different spot configurations could also explain the observed photometry. However, as a configuration exists that is compatible with the TESS photometry and the NEID radial velocities, we conclude that the observed RV slope is likely due to stellar activity.

\subsection{Obliquities as a Function of Age}
\begin{figure*}[t!]
\begin{center}
\includegraphics[width=0.9\textwidth]{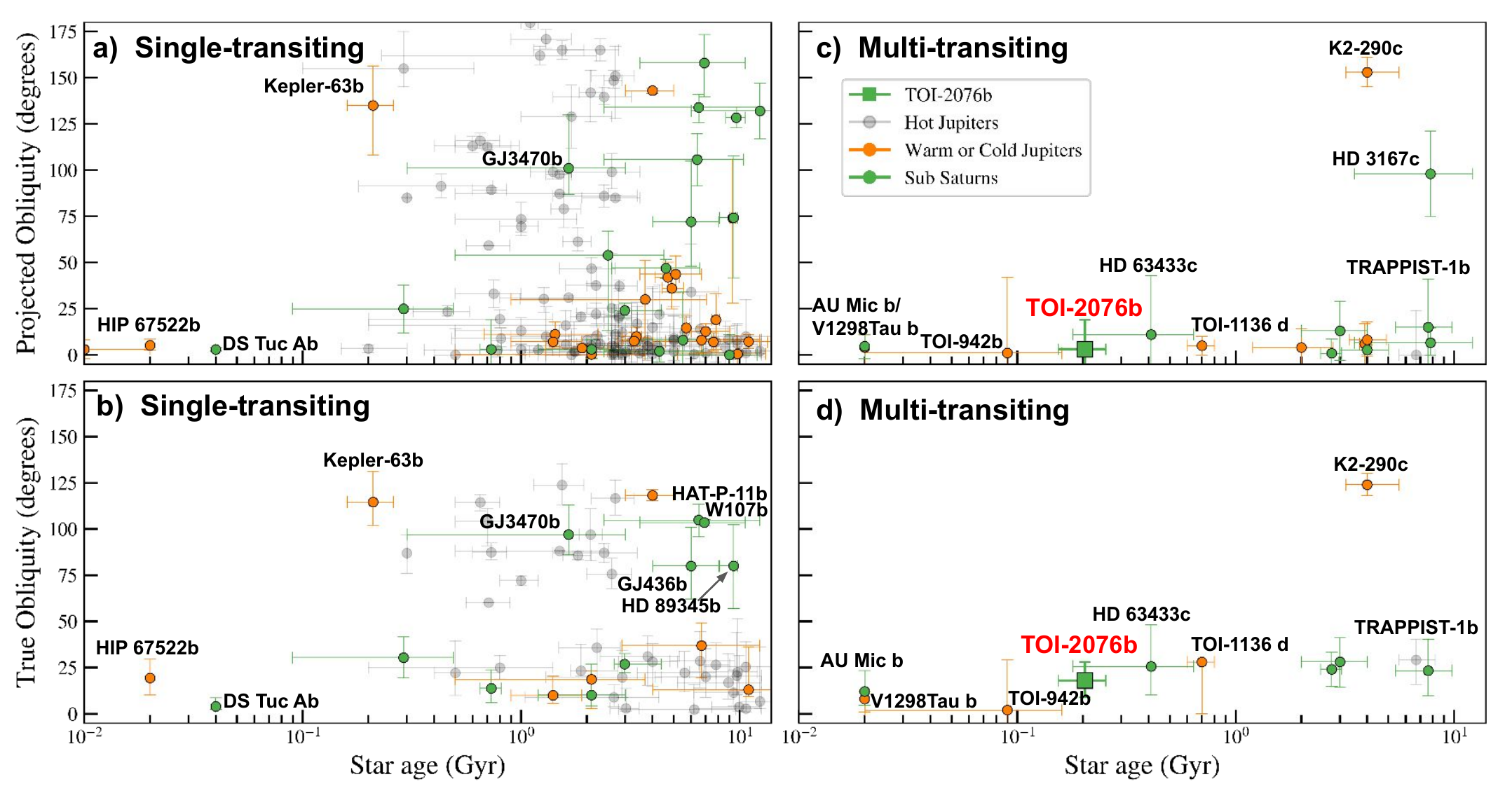}
\vspace{-0.4cm}
\end{center}
\caption{Sky-projected obliquities (a, c) and 3D obliquities (b, d) for planetary systems as a function of age for single-transiting (a, b), and multi-transiting (c, d) systems. Planets with masses $>0.3 \unit{M_{J}}$ and $a/R_\star < 10$ are classified hot Jupiters (black points), and as warm Jupiters if $a/R_\star>10$ (orange points). Planets with masses $<0.3 \unit{M_{J}}$ are classified as sub-Saturns regardless of the value of $a/R_\star$. This classification system is adopted from \cite{albrecht2022}. Despite a lack of a mass measurement of TOI-2076 b we classify it as a sub-Saturn due to its radius and distance. The position of TOI-2076 b is highlighted in red. Obliquity measurements for systems excluding TOI-2076 are drawn from \cite{albrecht2022}, \cite{dai2022}, \cite{bourrier2023}, and the TEPCAT database \citep{southworth2011} where the error on the sky-projected obliquity was $\Delta \lambda < 40^\circ$, and the fractional error on the age of the system was less than 90\%. W107b denotes WASP-107b. }
\label{fig:age}
\end{figure*}

TOI-2076 b joins a small but growing group of planets in young systems with multiple transiting planets with measured obliquities. Figure \ref{fig:age} compares the obliquity of TOI-2076 b to obliquity measurements of other known exoplanets in single and multi transiting planet systems as a function of age of the system. 

For systems with a single transiting planet, we see that highly misaligned planets are only seen for ages $\gtrsim$200 Myr, where the Kepler-63b system---hosting a Saturn-size planet in a 9.4 day orbit around a young 210MYr Sun-like star---is the youngest system hosting a highly misaligned planet with $\lambda=-135_{-26.8}^{+21.2\:\circ}$, and $\psi=114.6_{-12.5}^{+16.6\:\circ}$ \citep{sanchisojeda2013,bourrier2023}. For older systems ($\gtrsim$1 Gyr), there is a growing population of eccentric sub-Saturns that are on misaligned orbits, including WASP-107b \citep{rubenzahl2021}, GJ 436 b \citep{bourrier2018,bourrier2022}, GJ 3470b \citep{stefansson2022}, HD 89345b \citep{bourrier2023}, and HAT-P-11b (\cite{hirano2011hatp11} and \cite{winn2010}). Three of these systems---WASP-107b, GJ 3470b, and HAT-P-11b---have known outer companions or candidate outer companions that have been suggested as possible paths to explain the misalignments of the inner transiting planets via gravitational interactions. However, the story is not fully clear, as the GJ 436b system does not have a known outer companion in the system, and there are some sub-Saturns on eccentric orbits that are observed to be on well-aligned orbits \citep[e.g., K2-25b;][]{stefansson2020b}. This could possibly indicate that different formation mechanisms are at play. The fact that misaligned planets with high obliquities are only seen around systems with ages $\gtrsim$200 Myr, potentially points to that the origin of misalignment might not be primordial and is rather caused by dynamical interactions later on. However, the mechanisms that are often invoked to explain misalignments---secular resonance crossings due to a disappearing disk and a massive outer companion \citep{petrovich2020}, and Von Zeipel-Kozai-Lidov oscillations \citep{fabrycky2007,naoz2016}---occur on relatively fast timescales of $10^5-10^6$yrs, so the lack of misalignments of the youngest small planets continues to be noteworthy. Additional obliquity constraints of the very youngest systems ($<100$MYr) will be particularly valuable. 

For the systems with multiple transiting planets (multi-transiting systems), we see from Figure \ref{fig:age} that there are no known young multi-transiting systems with ages $\lesssim1-3$ Gyr on misaligned orbits. We see that there are two older misaligned multi-transiting systems: K2-290 \citep{hjorth2021}, and HD 3167c \citep{dalal2019}, with ages $\gtrsim$3 GYr. Interestingly, these systems have contrasting architectures: K2-290 hosts two planets, K2-290b and c in a co-planar orbit, whereas HD 3167b and c have a mutual inclinations of $\sim$90 degrees \citep{bourrier2021}. To explain the misalignments, for the K2-290 system it has been suggested that an outer star K2-290 B (projected separation of 110 au) could have tilted the protoplanetary disk of K2-290 A, causing K2-290b and c to form coplanar in an initially misaligned disk \citep{hjorth2021}. An alternative formation scenario was suggested by \cite{best2022}, in which the third star in the triple system, K2-290 C (projected separation of 2500 au), could be responsible for the misalignment of both planets through gravitational perturbations at much longer timescales (typically $\gtrsim 100$ Myr). For the HD 3167 system, \cite{bourrier2021} suggest the that the perpendicular architecture likely arose from the outer planet being tilted through gravitational interactions with a possible outer companion, while the inner ultra-short period planet likely retained a low obliquity due to tight tidal-coupling with the host star.

In contrast to the misaligned multi-planet systems, for TOI-2076 b there are multiple lines of evidence suggesting that TOI-2076 b formed via a more dynamically benign process of smooth disk migration in an initially well-aligned disk. First, the large $a/R_\star \sim 25$ value for TOI-2076 b makes tidal realignment inefficient, making TOI-2076 b a pristine probe of the initial formation angle. This, combined with the currently observed low-obliquity of TOI-2076 b, disfavors a scenario where TOI-2076 b experienced a high degree of misalignment that was subsequently realigned. Second, the planets in the TOI-2076 system orbit close to period commensurabilities (b and c at close to 2:1 resonance; c and d close to 5:3 resonance) with clear TTVs observed in the system \citep{osborn2022}, which demonstrates that the planets in the TOI-2076 system are tightly gravitationally interacting. Therefore, an appealing formation scenario for the compact TOI-2076 b system is through smooth disk migration within an initially well-aligned protoplanetary disk \citep[e.g.,][]{goldreich1979,goldreich1980}, where the planets migrated into their resonant orbits that we see today. Future RM effect observations of TOI-2076 c and d will help constrain the coplanarity of the system, which through these lines of evidence we would expect to be likely well-aligned with the orbit of planet b. TOI-2076b is similar to the recently studied TOI-1136 system \citep{dai2022}, which is a compact network of at least 6 transiting planets in a resonant chain that likely formed through smooth disk migration in an initially well-aligned disk.

\section{Summary}
\label{sec:summary}
Using high precision in-transit spectroscopic observations with the NEID spectrograph on the WIYN 3.5m Telescope at Kitt Peak Observatory, we determined that the young (204 $\pm 50$ Myr) sub-Neptune planet TOI-2076 b has a low sky-projected obliquity of \reslambda. Leveraging knowledge of the size of the star, and the stellar rotation period, we estimate an obliquity of \respsi\ and a stellar inclination of \resistar\ suggesting that TOI-2076 b is on an orbit well-aligned with the stellar equator of its host star. Three sectors of data from the TESS spacecraft along with precise diffuser-assisted photometry from ARCTIC on the ARC 3.5 m telescope at APO were used to precisely constrain the orbital parameters of the planet.

TOI-2076 b joins a small, but growing sample of young multi-planet systems on well-aligned orbits. It is the fourth planet with an age $\leq 300$ Myr in a multi-transiting system with an obliquity measurement. The well-aligned orbit of TOI-2076 b together with the compact multi-planet configuration that shows evidence of transit timing variations suggests that the TOI-2076 system likely formed via convergent disk-migration in an initially well-aligned disk. This would suggest that TOI-2076c and d are likely coplanar to TOI-2076 b. Additional RM observations of the outer planets are needed to confirm this hypothesis. 

In addition, due to its brightness, TOI-2076 hosts some of the most accessible young planets for atmospheric characterization. With a measurement of its obliquity, the architecture of the TOI-2076 system is now better understood, which will help place any future follow-up observations---such as atmospheric characterization---in further context.

\facilities{NEID/WIYN 3.5m, ARCTIC/ARC 3.5m, \textit{Gaia}.} 
\software{AstroImageJ \citep{collins2017}, 
\texttt{astroplan} \citep{morris2018},
\texttt{astropy} \citep{astropy2013},
\texttt{astroscrappy} \citep{astroscrappy}
\texttt{astroquery} \citep{astroquery},
\texttt{barycorrpy} \citep{kanodia2018}, 
\texttt{batman} \citep{kreidberg2015},
\texttt{celerite} \citep{Foreman-Mackey2017}, 
\texttt{corner.py} \citep{dfm2016}, 
\texttt{dynesty} \citep{speagle2019}, 
\texttt{emcee} \citep{dfm2013},
\texttt{iDiffuse} \citep{stefansson2018b},
\texttt{juliet} \citep{Espinoza2019},
\texttt{Jupyter} \citep{jupyter2016},
\texttt{lightkurve} \cite{lightkurve},
\texttt{matplotlib} \citep{hunter2007},
\texttt{numpy} \citep{vanderwalt2011},
\texttt{pandas} \citep{pandas2010},
\texttt{pyde} \citep{pyde},
\texttt{radvel} \citep{fulton2018},
\texttt{rmfit} \citep{stefansson2022},
\texttt{SOAP2.0} \citep{dumusque2014},
\texttt{SERVAL} \citep{zechmeister2018}.}

\acknowledgements

GS acknowledges support provided by NASA through the NASA Hubble Fellowship grant HST-HF2-51519.001-A awarded by the Space Telescope Science Institute, which is operated by the Association of Universities for Research in Astronomy, Inc., for NASA, under contract NAS5-26555. GS acknowledges support through the Henry Norris Russell Fellowship at Princeton during the preparation of this manuscript.

Data presented were obtained by the NEID spectrograph built by Penn State University and operated at the WIYN Observatory by NOIRLab, under the NN-EXPLORE partnership of the National Aeronautics and Space Administration and the National Science Foundation. Based in part on observations at the Kitt Peak National Observatory, NSF’s NOIRLab (Prop. ID 2022A-970114; PI: G. Stefansson), managed by the Association of Universities for Research in Astronomy (AURA) under a cooperative agreement with the National Science Foundation. WIYN is a joint facility of the University of Wisconsin–Madison, Indiana University, NSF’s NOIRLab, the Pennsylvania State University, Purdue University, University of California, Irvine, and the University of Missouri. The authors are honored to be permitted to conduct astronomical research on Iolkam Du’ag (Kitt Peak), a mountain with particular significance to the Tohono O’odham. Data presented herein were obtained at the WIYN Observatory from telescope time allocated to NN-EXPLORE through the scientific partnership of the National Aeronautics and Space Administration, the National Science Foundation, and the National Optical Astronomy Observatory. This work was supported by a NASA WIYN PI Data Award, administered by the NASA Exoplanet Science Institute. We thank the NEID Queue Observers and WIYN Observing Associates for their skillful execution of our NEID observations.  We extend our deepest gratitude to Zade Arnold, Joe Davis, Michelle Edwards, John Ehret, Tina Juan, Brian Pisarek, Aaron Rowe, Fred Wortman, the Eastern Area Incident Management Team, and all of the firefighters and air support crew who fought the recent Contreras fire. Against great odds, you saved Kitt Peak National Observatory.

These results are based on observations obtained with the Apache Point Observatory 3.5-meter telescope which is owned and operated by the Astrophysical Research Consortium. We wish to thank the APO 3.5m telescope operators in their assistance in obtaining these data.

This work was partially supported by funding from the Center for Exoplanets and Habitable Worlds. The Center for Exoplanets and Habitable Worlds is supported by the Pennsylvania State University, the Eberly College of Science, and the Pennsylvania Space Grant Consortium. CIC acknowledges support by NASA Headquarters through an appointment to the NASA Postdoctoral Program at the Goddard Space Flight Center, administered by URSA through a contract with NASA. This work was performed for the Jet Propulsion Laboratory, California Institute of Technology, sponsored by the United States Government under the Prime Contract 80NM0018D0004 between Caltech and NASA. We acknowledge support from NSF grant AST-1909506, AST-190950, AST-1910954, AST-1907622, and AST-1907622, and the Research Corporation for precision photometric observations with diffuser-assisted photometry. Computations for this research were performed on the Pennsylvania State University’s Institute for Computational \& Data Sciences (ICDS). A portion of this work was enabled by support from the Mt Cuba Astronomical Foundation. C.P. acknowledges support from ANID Millennium Science Initiative-ICN12\_009, CATA-Basal AFB-170002, ANID BASAL project FB210003, FONDECYT Regular grant 1210425 and ANID+REC Convocatoria Nacional subvencion a la instalacion en la Academia convocatoria 2020 PAI77200076. LD and HMC acknowledge funding from a UKRI Future Leader Fellowship, grant number MR/S035214/1.

This work has made use of data from the European Space Agency (ESA) mission Gaia processed by the Gaia Data Processing and Analysis Consortium (DPAC). Funding for the DPAC has been provided by national institutions, in particular the institutions participating in the Gaia Multilateral Agreement.

This research made use of the NASA Exoplanet Archive, which is operated by the California Institute of Technology, under contract with the National Aeronautics and Space Administration under the Exoplanet Exploration Program. This research made use of Astropy, a community-developed core Python package for Astronomy \citep{astropy2013}.

\bibliography{references}{}

\begin{thebibliography}{}
\expandafter\ifx\csname natexlab\endcsname\relax\def\natexlab#1{#1}\fi
\providecommand{\url}[1]{\href{#1}{#1}}
\providecommand{\dodoi}[1]{doi:~\href{http://doi.org/#1}{\nolinkurl{#1}}}
\providecommand{\doeprint}[1]{\href{http://ascl.net/#1}{\nolinkurl{http://ascl.net/#1}}}
\providecommand{\doarXiv}[1]{\href{https://arxiv.org/abs/#1}{\nolinkurl{https://arxiv.org/abs/#1}}}

\bibitem[{{Albrecht} {et~al.}(2012){Albrecht}, {Winn}, {Johnson}, {Howard},
  {Marcy}, {Butler}, {Arriagada}, {Crane}, {Shectman}, {Thompson}, {Hirano},
  {Bakos}, \& {Hartman}}]{albrecht2012}
{Albrecht}, S., {Winn}, J.~N., {Johnson}, J.~A., {et~al.} 2012, \apj, 757, 18,
  \dodoi{10.1088/0004-637X/757/1/18}

\bibitem[{{Albrecht} {et~al.}(2022){Albrecht}, {Dawson}, \&
  {Winn}}]{albrecht2022}
{Albrecht}, S.~H., {Dawson}, R.~I., \& {Winn}, J.~N. 2022, \pasp, 134, 082001,
  \dodoi{10.1088/1538-3873/ac6c09}

\bibitem[{{Albrecht} {et~al.}(2021){Albrecht}, {Marcussen}, {Winn}, {Dawson},
  \& {Knudstrup}}]{albrecht2021}
{Albrecht}, S.~H., {Marcussen}, M.~L., {Winn}, J.~N., {Dawson}, R.~I., \&
  {Knudstrup}, E. 2021, \apjl, 916, L1, \dodoi{10.3847/2041-8213/ac0f03}

\bibitem[{{Astropy Collaboration} {et~al.}(2013){Astropy Collaboration},
  {Robitaille}, {Tollerud}, {Greenfield}, {Droettboom}, {Bray}, {Aldcroft},
  {Davis}, {Ginsburg}, {Price-Whelan}, {Kerzendorf}, {Conley}, {Crighton},
  {Barbary}, {Muna}, {Ferguson}, {Grollier}, {Parikh}, {Nair}, {Unther},
  {Deil}, {Woillez}, {Conseil}, {Kramer}, {Turner}, {Singer}, {Fox}, {Weaver},
  {Zabalza}, {Edwards}, {Azalee Bostroem}, {Burke}, {Casey}, {Crawford},
  {Dencheva}, {Ely}, {Jenness}, {Labrie}, {Lim}, {Pierfederici}, {Pontzen},
  {Ptak}, {Refsdal}, {Servillat}, \& {Streicher}}]{astropy2013}
{Astropy Collaboration}, {Robitaille}, T.~P., {Tollerud}, E.~J., {et~al.} 2013,
  \aap, 558, A33, \dodoi{10.1051/0004-6361/201322068}

\bibitem[{{Benatti} {et~al.}(2019){Benatti}, {Nardiello}, {Malavolta},
  {Desidera}, {Borsato}, {Nascimbeni}, {Damasso}, {D'Orazi}, {Mesa}, {Messina},
  {Esposito}, {Bignamini}, {Claudi}, {Covino}, {Lovis}, \&
  {Sabotta}}]{benatti2019}
{Benatti}, S., {Nardiello}, D., {Malavolta}, L., {et~al.} 2019, \aap, 630, A81,
  \dodoi{10.1051/0004-6361/201935598}

\bibitem[{{Benz} {et~al.}(2021){Benz}, {Broeg}, {Fortier}, {Rando}, {Beck},
  {Beck}, {Queloz}, {Ehrenreich}, {Maxted}, {Isaak}, {Billot}, {Alibert},
  {Alonso}, {Ant{\'o}nio}, {Asquier}, {Bandy}, {B{\'a}rczy}, {Barrado},
  {Barros}, {Baumjohann}, {Bekkelien}, {Bergomi}, {Biondi}, {Bonfils},
  {Borsato}, {Brandeker}, {Busch}, {Cabrera}, {Cessa}, {Charnoz}, {Chazelas},
  {Collier Cameron}, {Corral Van Damme}, {Cortes}, {Davies}, {Deleuil},
  {Deline}, {Delrez}, {Demangeon}, {Demory}, {Erikson}, {Farinato}, {Fossati},
  {Fridlund}, {Futyan}, {Gandolfi}, {Garcia Munoz}, {Gillon}, {Guterman},
  {Gutierrez}, {Hasiba}, {Heng}, {Hernandez}, {Hoyer}, {Kiss}, {Kovacs},
  {Kuntzer}, {Laskar}, {Lecavelier des Etangs}, {Lendl}, {L{\'o}pez}, {Lora},
  {Lovis}, {L{\"u}ftinger}, {Magrin}, {Malvasio}, {Marafatto}, {Michaelis}, {de
  Miguel}, {Modrego}, {Munari}, {Nascimbeni}, {Olofsson}, {Ottacher},
  {Ottensamer}, {Pagano}, {Palacios}, {Pall{\'e}}, {Peter}, {Piazza}, {Piotto},
  {Pizarro}, {Pollaco}, {Ragazzoni}, {Ratti}, {Rauer}, {Ribas}, {Rieder},
  {Rohlfs}, {Safa}, {Salatti}, {Santos}, {Scandariato}, {S{\'e}gransan},
  {Simon}, {Smith}, {Sordet}, {Sousa}, {Steller}, {Szab{\'o}}, {Szoke},
  {Thomas}, {Tschentscher}, {Udry}, {Van Grootel}, {Viotto}, {Walter},
  {Walton}, {Wildi}, \& {Wolter}}]{benz2021}
{Benz}, W., {Broeg}, C., {Fortier}, A., {et~al.} 2021, Experimental Astronomy,
  51, 109, \dodoi{10.1007/s10686-020-09679-4}

\bibitem[{{Best} \& {Petrovich}(2022)}]{best2022}
{Best}, S., \& {Petrovich}, C. 2022, \apjl, 925, L5,
  \dodoi{10.3847/2041-8213/ac49e9}

\bibitem[{{Bourrier} {et~al.}(2021){Bourrier}, {Lovis, C.}, {Cretignier, M.},
  {Allart, R.}, {Dumusque, X.}, {Delisle, J.-B.}, {Deline, A.}, {Sousa, S. G.},
  {Adibekyan, V.}, {Alibert, Y.}, {Barros, S. C. C.}, {Borsa, F.}, {Cristiani,
  S.}, {Demangeon, O.}, {Ehrenreich, D.}, {Figueira, P.}, {Gonz\'alez
  Hern\'andez, J. I.}, {Lendl, M.}, {Lillo-Box, J.}, {Lo Curto, G.}, {Di
  Marcantonio, P.}, {Martins, C. J. A. P.}, {M\'egevand, D.}, {Mehner, A.},
  {Micela, G.}, {Molaro, P.}, {Oshagh, M.}, {Palle, E.}, {Pepe, F.}, {Poretti,
  E.}, {Rebolo, R.}, {Santos, N. C.}, {Scandariato, G.}, {Seidel, J. V.},
  {Sozzetti, A.}, {Su\'arez Mascare\~no, A.}, \& {Zapatero Osorio, M.
  R.}}]{bourrier2021}
{Bourrier}, {Lovis, C.}, {Cretignier, M.}, {et~al.} 2021, A\&A, 654, A152,
  \dodoi{10.1051/0004-6361/202141527}

\bibitem[{{Bourrier} {et~al.}(2018){Bourrier}, {Lovis}, {Beust}, {Ehrenreich},
  {Henry}, {Astudillo-Defru}, {Allart}, {Bonfils}, {S{\'e}gransan}, {Delfosse},
  {Cegla}, {Wyttenbach}, {Heng}, {Lavie}, \& {Pepe}}]{bourrier2018}
{Bourrier}, V., {Lovis}, C., {Beust}, H., {et~al.} 2018, \nat, 553, 477,
  \dodoi{10.1038/nature24677}

\bibitem[{{Bourrier} {et~al.}(2022){Bourrier}, {Zapatero Osorio}, {Allart},
  {Attia}, {Cretignier}, {Dumusque}, {Lovis}, {Adibekyan}, {Borsa}, {Figueira},
  {Gonz{\'a}lez Hern{\'a}ndez}, {Mehner}, {Santos}, {Schmidt}, {Seidel},
  {Sozzetti}, {Alibert}, {Casasayas-Barris}, {Ehrenreich}, {Lo Curto},
  {Martins}, {Di Marcantonio}, {M{\'e}gevand}, {Nunes}, {Palle}, {Poretti}, \&
  {Sousa}}]{bourrier2022}
{Bourrier}, V., {Zapatero Osorio}, M.~R., {Allart}, R., {et~al.} 2022, \aap,
  663, A160, \dodoi{10.1051/0004-6361/202142559}

\bibitem[{{Bourrier} {et~al.}(2023){Bourrier}, {Attia}, {Mallonn}, {Marret},
  {Lendl}, {Konig}, {Krenn}, {Cretignier}, {Allart}, {Henry}, {Bryant},
  {Leleu}, {Nielsen}, {Hebrard}, {Hara}, {Ehrenreich}, {Seidel}, {dos Santos},
  {Lovis}, {Bayliss}, {Cegla}, {Dumusque}, {Boisse}, {Boucher}, {Bouchy},
  {Pepe}, {Lavie}, {Rey Cerda}, {S{\'e}gransan}, {Udry}, \&
  {Vrignaud}}]{bourrier2023}
{Bourrier}, V., {Attia}, O., {Mallonn}, M., {et~al.} 2023, \aap, 669, A63,
  \dodoi{10.1051/0004-6361/202245004}

\bibitem[{{Chen} \& {Kipping}(2017)}]{chen2017}
{Chen}, J., \& {Kipping}, D. 2017, \apj, 834, 17,
  \dodoi{10.3847/1538-4357/834/1/17}

\bibitem[{{Coelho} {et~al.}(2005){Coelho}, {Barbuy}, {Mel{\'e}ndez},
  {Schiavon}, \& {Castilho}}]{coelho2005}
{Coelho}, P., {Barbuy}, B., {Mel{\'e}ndez}, J., {Schiavon}, R.~P., \&
  {Castilho}, B.~V. 2005, \aap, 443, 735, \dodoi{10.1051/0004-6361:20053511}

\bibitem[{{Collins} {et~al.}(2017){Collins}, {Kielkopf}, {Stassun}, \&
  {Hessman}}]{collins2017}
{Collins}, K.~A., {Kielkopf}, J.~F., {Stassun}, K.~G., \& {Hessman}, F.~V.
  2017, \aj, 153, 77, \dodoi{10.3847/1538-3881/153/2/77}

\bibitem[{{Dai} {et~al.}(2023){Dai}, {Masuda}, {Beard}, {Robertson},
  {Goldberg}, {Batygin}, {Bouma}, {Lissauer}, {Knudstrup}, {Albrecht},
  {Howard}, {Knutson}, {Petigura}, {Weiss}, {Isaacson}, {Kristiansen},
  {Osborn}, {Wang}, {Wang}, {Behmard}, {Greklek-McKeon}, {Vissapragada},
  {Batalha}, {Brinkman}, {Chontos}, {Crossfield}, {Dressing}, {Fetherolf},
  {Fulton}, {Hill}, {Huber}, {Kane}, {Lubin}, {MacDougall}, {Mayo},
  {Mo{\v{c}}nik}, {Akana Murphy}, {Rubenzahl}, {Scarsdale}, {Tyler}, {Zandt},
  {Polanski}, {Schwengeler}, {Terentev}, {Benni}, {Bieryla}, {Ciardi}, {Falk},
  {Furlan}, {Girardin}, {Guerra}, {Hesse}, {Howell}, {Lillo-Box}, {Matthews},
  {Twicken}, {Villase{\~n}or}, {Latham}, {Jenkins}, {Ricker}, {Seager},
  {Vanderspek}, \& {Winn}}]{dai2022}
{Dai}, F., {Masuda}, K., {Beard}, C., {et~al.} 2023, \aj, 165, 33,
  \dodoi{10.3847/1538-3881/aca327}

\bibitem[{{Dalal} {et~al.}(2019){Dalal}, {H{\'e}brard}, {Lecavelier des
  {\'E}tangs}, {Petit}, {Bourrier}, {Laskar}, {K{\"o}nig}, \&
  {Correia}}]{dalal2019}
{Dalal}, S., {H{\'e}brard}, G., {Lecavelier des {\'E}tangs}, A., {et~al.} 2019,
  \aap, 631, A28, \dodoi{10.1051/0004-6361/201935944}

\bibitem[{{Dumusque} {et~al.}(2014){Dumusque}, {Boisse}, \&
  {Santos}}]{dumusque2014}
{Dumusque}, X., {Boisse}, I., \& {Santos}, N.~C. 2014, \apj, 796, 132,
  \dodoi{10.1088/0004-637X/796/2/132}

\bibitem[{{Eastman} {et~al.}(2019){Eastman}, {Rodriguez}, {Agol}, {Stassun},
  {Beatty}, {Vanderburg}, {Gaudi}, {Collins}, \& {Luger}}]{eastman2019}
{Eastman}, J.~D., {Rodriguez}, J.~E., {Agol}, E., {et~al.} 2019, arXiv e-prints
  (submitted to PASP).
\newblock \doarXiv{1907.09480}

\bibitem[{{Espinoza} {et~al.}(2019{\natexlab{a}}){Espinoza}, {Kossakowski}, \&
  {Brahm}}]{juliet}
{Espinoza}, N., {Kossakowski}, D., \& {Brahm}, R. 2019{\natexlab{a}}, \mnras,
  490, 2262, \dodoi{10.1093/mnras/stz2688}

\bibitem[{{Espinoza} {et~al.}(2019{\natexlab{b}}){Espinoza}, {Kossakowski}, \&
  {Brahm}}]{Espinoza2019}
---. 2019{\natexlab{b}}, \mnras, 490, 2262, \dodoi{10.1093/mnras/stz2688}

\bibitem[{{Fabrycky} \& {Tremaine}(2007)}]{fabrycky2007}
{Fabrycky}, D., \& {Tremaine}, S. 2007, \apj, 669, 1298, \dodoi{10.1086/521702}

\bibitem[{Foreman-Mackey(2016)}]{dfm2016}
Foreman-Mackey, D. 2016, JOSS, 24, \dodoi{10.21105/joss.00024}

\bibitem[{{Foreman-Mackey} {et~al.}(2017){Foreman-Mackey}, {Agol},
  {Ambikasaran}, \& {Angus}}]{Foreman-Mackey2017}
{Foreman-Mackey}, D., {Agol}, E., {Ambikasaran}, S., \& {Angus}, R. 2017, \aj,
  154, 220, \dodoi{10.3847/1538-3881/aa9332}

\bibitem[{{Foreman-Mackey} {et~al.}(2013){Foreman-Mackey}, {Hogg}, {Lang}, \&
  {Goodman}}]{dfm2013}
{Foreman-Mackey}, D., {Hogg}, D.~W., {Lang}, D., \& {Goodman}, J. 2013, PASP,
  125, 306, \dodoi{10.1086/670067}

\bibitem[{{Fulton} {et~al.}(2018){Fulton}, {Petigura}, {Blunt}, \&
  {Sinukoff}}]{fulton2018}
{Fulton}, B.~J., {Petigura}, E.~A., {Blunt}, S., \& {Sinukoff}, E. 2018, \pasp,
  130, 044504, \dodoi{10.1088/1538-3873/aaaaa8}

\bibitem[{{Gaia Collaboration} {et~al.}(2021){Gaia Collaboration}, {Brown},
  {Vallenari}, {Prusti}, {de Bruijne}, {Babusiaux}, {Biermann}, {Creevey},
  {Evans}, {Eyer}, {Hutton}, {Jansen}, {Jordi}, {Klioner}, {Lammers},
  {Lindegren}, {Luri}, {Mignard}, {Panem}, {Pourbaix}, {Randich}, {Sartoretti},
  {Soubiran}, {Walton}, {Arenou}, {Bailer-Jones}, {Bastian}, {Cropper},
  {Drimmel}, {Katz}, {Lattanzi}, {van Leeuwen}, {Bakker}, {Cacciari},
  {Casta{\~n}eda}, {De Angeli}, {Ducourant}, {Fabricius}, {Fouesneau},
  {Fr{\'e}mat}, {Guerra}, {Guerrier}, {Guiraud}, {Jean-Antoine Piccolo},
  {Masana}, {Messineo}, {Mowlavi}, {Nicolas}, {Nienartowicz}, {Pailler},
  {Panuzzo}, {Riclet}, {Roux}, {Seabroke}, {Sordo}, {Tanga}, {Th{\'e}venin},
  {Gracia-Abril}, {Portell}, {Teyssier}, {Altmann}, {Andrae}, {Bellas-Velidis},
  {Benson}, {Berthier}, {Blomme}, {Brugaletta}, {Burgess}, {Busso}, {Carry},
  {Cellino}, {Cheek}, {Clementini}, {Damerdji}, {Davidson}, {Delchambre},
  {Dell'Oro}, {Fern{\'a}ndez-Hern{\'a}ndez}, {Galluccio}, {Garc{\'\i}a-Lario},
  {Garcia-Reinaldos}, {Gonz{\'a}lez-N{\'u}{\~n}ez}, {Gosset}, {Haigron},
  {Halbwachs}, {Hambly}, {Harrison}, {Hatzidimitriou}, {Heiter},
  {Hern{\'a}ndez}, {Hestroffer}, {Hodgkin}, {Holl}, {Jan{\ss}en}, {Jevardat de
  Fombelle}, {Jordan}, {Krone-Martins}, {Lanzafame}, {L{\"o}ffler}, {Lorca},
  {Manteiga}, {Marchal}, {Marrese}, {Moitinho}, {Mora}, {Muinonen}, {Osborne},
  {Pancino}, {Pauwels}, {Petit}, {Recio-Blanco}, {Richards}, {Riello},
  {Rimoldini}, {Robin}, {Roegiers}, {Rybizki}, {Sarro}, {Siopis}, {Smith},
  {Sozzetti}, {Ulla}, {Utrilla}, {van Leeuwen}, {van Reeven}, {Abbas}, {Abreu
  Aramburu}, {Accart}, {Aerts}, {Aguado}, {Ajaj}, {Altavilla}, {{\'A}lvarez},
  {{\'A}lvarez Cid-Fuentes}, {Alves}, {Anderson}, {Anglada Varela}, {Antoja},
  {Audard}, {Baines}, {Baker}, {Balaguer-N{\'u}{\~n}ez}, {Balbinot}, {Balog},
  {Barache}, {Barbato}, {Barros}, {Barstow}, {Bartolom{\'e}}, {Bassilana},
  {Bauchet}, {Baudesson-Stella}, {Becciani}, {Bellazzini}, {Bernet}, {Bertone},
  {Bianchi}, {Blanco-Cuaresma}, {Boch}, {Bombrun}, {Bossini}, {Bouquillon},
  {Bragaglia}, {Bramante}, {Breedt}, {Bressan}, {Brouillet}, {Bucciarelli},
  {Burlacu}, {Busonero}, {Butkevich}, {Buzzi}, {Caffau}, {Cancelliere},
  {C{\'a}novas}, {Cantat-Gaudin}, {Carballo}, {Carlucci}, {Carnerero},
  {Carrasco}, {Casamiquela}, {Castellani}, {Castro-Ginard}, {Castro Sampol},
  {Chaoul}, {Charlot}, {Chemin}, {Chiavassa}, {Cioni}, {Comoretto}, {Cooper},
  {Cornez}, {Cowell}, {Crifo}, {Crosta}, {Crowley}, {Dafonte}, {Dapergolas},
  {David}, {David}, {de Laverny}, {De Luise}, {De March}, {De Ridder}, {de
  Souza}, {de Teodoro}, {de Torres}, {del Peloso}, {del Pozo}, {Delbo},
  {Delgado}, {Delgado}, {Delisle}, {Di Matteo}, {Diakite}, {Diener},
  {Distefano}, {Dolding}, {Eappachen}, {Edvardsson}, {Enke}, {Esquej}, {Fabre},
  {Fabrizio}, {Faigler}, {Fedorets}, {Fernique}, {Fienga}, {Figueras},
  {Fouron}, {Fragkoudi}, {Fraile}, {Franke}, {Gai}, {Garabato},
  {Garcia-Gutierrez}, {Garc{\'\i}a-Torres}, {Garofalo}, {Gavras}, {Gerlach},
  {Geyer}, {Giacobbe}, {Gilmore}, {Girona}, {Giuffrida}, {Gomel}, {Gomez},
  {Gonzalez-Santamaria}, {Gonz{\'a}lez-Vidal}, {Granvik},
  {Guti{\'e}rrez-S{\'a}nchez}, {Guy}, {Hauser}, {Haywood}, {Helmi}, {Hidalgo},
  {Hilger}, {H{\l}adczuk}, {Hobbs}, {Holland}, {Huckle}, {Jasniewicz},
  {Jonker}, {Juaristi Campillo}, {Julbe}, {Karbevska}, {Kervella}, {Khanna},
  {Kochoska}, {Kontizas}, {Kordopatis}, {Korn}, {Kostrzewa-Rutkowska},
  {Kruszy{\'n}ska}, {Lambert}, {Lanza}, {Lasne}, {Le Campion}, {Le Fustec},
  {Lebreton}, {Lebzelter}, {Leccia}, {Leclerc}, {Lecoeur-Taibi}, {Liao},
  {Licata}, {Lindstr{\o}m}, {Lister}, {Livanou}, {Lobel}, {Madrero Pardo},
  {Managau}, {Mann}, {Marchant}, {Marconi}, {Marcos Santos}, {Marinoni},
  {Marocco}, {Marshall}, {Martin Polo}, {Mart{\'\i}n-Fleitas}, {Masip},
  {Massari}, {Mastrobuono-Battisti}, {Mazeh}, {McMillan}, {Messina},
  {Michalik}, {Millar}, {Mints}, {Molina}, {Molinaro}, {Moln{\'a}r},
  {Montegriffo}, {Mor}, {Morbidelli}, {Morel}, {Morris}, {Mulone}, {Munoz},
  {Muraveva}, {Murphy}, {Musella}, {Noval}, {Ord{\'e}novic}, {Orr{\`u}},
  {Osinde}, {Pagani}, {Pagano}, {Palaversa}, {Palicio}, {Panahi}, {Pawlak},
  {Pe{\~n}alosa Esteller}, {Penttil{\"a}}, {Piersimoni}, {Pineau}, {Plachy},
  {Plum}, {Poggio}, {Poretti}, {Poujoulet}, {Pr{\v{s}}a}, {Pulone}, {Racero},
  {Ragaini}, {Rainer}, {Raiteri}, {Rambaux}, {Ramos}, {Ramos-Lerate}, {Re
  Fiorentin}, {Regibo}, {Reyl{\'e}}, {Ripepi}, {Riva}, {Rixon}, {Robichon},
  {Robin}, {Roelens}, {Rohrbasser}, {Romero-G{\'o}mez}, {Rowell}, {Royer},
  {Rybicki}, {Sadowski}, {Sagrist{\`a} Sell{\'e}s}, {Sahlmann}, {Salgado},
  {Salguero}, {Samaras}, {Sanchez Gimenez}, {Sanna}, {Santove{\~n}a},
  {Sarasso}, {Schultheis}, {Sciacca}, {Segol}, {Segovia}, {S{\'e}gransan},
  {Semeux}, {Shahaf}, {Siddiqui}, {Siebert}, {Siltala}, {Slezak}, {Smart},
  {Solano}, {Solitro}, {Souami}, {Souchay}, {Spagna}, {Spoto}, {Steele},
  {Steidelm{\"u}ller}, {Stephenson}, {S{\"u}veges}, {Szabados}, {Szegedi-Elek},
  {Taris}, {Tauran}, {Taylor}, {Teixeira}, {Thuillot}, {Tonello}, {Torra},
  {Torra}, {Turon}, {Unger}, {Vaillant}, {van Dillen}, {Vanel}, {Vecchiato},
  {Viala}, {Vicente}, {Voutsinas}, {Weiler}, {Wevers}, {Wyrzykowski}, {Yoldas},
  {Yvard}, {Zhao}, {Zorec}, {Zucker}, {Zurbach}, \& {Zwitter}}]{gaia2018}
{Gaia Collaboration}, {Brown}, A.~G.~A., {Vallenari}, A., {et~al.} 2021, \aap,
  649, A1, \dodoi{10.1051/0004-6361/202039657}

\bibitem[{{Gaidos} {et~al.}(2023){Gaidos}, {Hirano}, {Lee}, {Harakawa},
  {Hodapp}, {Jacobson}, {Kotani}, {Kudo}, {Kurokawa}, {Kuzuhara}, {Nishikawa},
  {Omiya}, {Serizawa}, {Tamura}, {Ueda}, \& {Vievard}}]{gaidos2022}
{Gaidos}, E., {Hirano}, T., {Lee}, R.~A., {et~al.} 2023, \mnras, 518, 3777,
  \dodoi{10.1093/mnras/stac3301}

\bibitem[{{Gaudi} \& {Winn}(2007)}]{gaudi2007}
{Gaudi}, B.~S., \& {Winn}, J.~N. 2007, \apj, 655, 550, \dodoi{10.1086/509910}

\bibitem[{Ginsburg {et~al.}(2018)Ginsburg, Sipocz, Parikh, Woillez, Groener,
  Liedtke, Robitaille, Deil, jcsegovia, Norman, Svoboda, Brasseur, Tollerud,
  Persson, adamginsburg, Séguin-Charbonneau, Armstrong, de~Val-Borro, Morris,
  Mirocha, Yadav, Seifert, Droettboom, Moolekamp, james allen, Bostroem,
  Egeland, Singer, Rol, \& Grollier}]{astroquery}
Ginsburg, A., Sipocz, B., Parikh, M., {et~al.} 2018, astropy/astroquery: v0.3.7
  release, \dodoi{10.5281/zenodo.1160627}

\bibitem[{{Goldreich} \& {Tremaine}(1979)}]{goldreich1979}
{Goldreich}, P., \& {Tremaine}, S. 1979, \apj, 233, 857, \dodoi{10.1086/157448}

\bibitem[{{Goldreich} \& {Tremaine}(1980)}]{goldreich1980}
---. 1980, \apj, 241, 425, \dodoi{10.1086/158356}

\bibitem[{{Halverson} {et~al.}(2016){Halverson}, {Terrien}, {Mahadevan}, {Roy},
  {Bender}, {Stef{\'a}nsson}, {Monson}, {Levi}, {Hearty}, {Blake}, {McElwain},
  {Schwab}, {Ramsey}, {Wright}, {Wang}, {Gong}, \& {Roberston}}]{halverson2016}
{Halverson}, S., {Terrien}, R., {Mahadevan}, S., {et~al.} 2016, in \procspie,
  Vol. 9908, Ground-based and Airborne Instrumentation for Astronomy VI,
  99086P, \dodoi{10.1117/12.2232761}

\bibitem[{{Hedges} {et~al.}(2021){Hedges}, {Hughes}, {Zhou}, {David}, {Becker},
  {Giacalone}, {Vanderburg}, {Rodriguez}, {Bieryla}, {Wirth}, {Atherton},
  {Fetherolf}, {Collins}, {Price-Whelan}, {Bedell}, {Quinn}, {Gan}, {Ricker},
  {Latham}, {Vanderspek}, {Seager}, {Winn}, {Jenkins}, {Kielkopf}, {Schwarz},
  {Dressing}, {Gonzales}, {Crossfield}, {Matthews}, {Jensen}, {Furlan},
  {Gnilka}, {Howell}, {Lester}, {Scott}, {Feliz}, {Lund}, {Siverd}, {Stevens},
  {Narita}, {Fukui}, {Murgas}, {Palle}, {Sutton}, {Stassun}, {Bouma}, {Vezie},
  {Villase{\~n}or}, {Quintana}, \& {Smith}}]{hedges2021}
{Hedges}, C., {Hughes}, A., {Zhou}, G., {et~al.} 2021, \aj, 162, 54,
  \dodoi{10.3847/1538-3881/ac06cd}

\bibitem[{{Hirano} {et~al.}(2011{\natexlab{a}}){Hirano}, {Narita}, {Shporer},
  {Sato}, {Aoki}, \& {Tamura}}]{hirano2011hatp11}
{Hirano}, T., {Narita}, N., {Shporer}, A., {et~al.} 2011{\natexlab{a}}, \pasj,
  63, 531, \dodoi{10.1093/pasj/63.sp2.S531}

\bibitem[{{Hirano} {et~al.}(2011{\natexlab{b}}){Hirano}, {Suto}, {Winn},
  {Taruya}, {Narita}, {Albrecht}, \& {Sato}}]{hirano2011}
{Hirano}, T., {Suto}, Y., {Winn}, J.~N., {et~al.} 2011{\natexlab{b}}, \apj,
  742, 69, \dodoi{10.1088/0004-637X/742/2/69}

\bibitem[{{Hjorth} {et~al.}(2021){Hjorth}, {Albrecht}, {Hirano}, {Winn},
  {Dawson}, {Zanazzi}, {Knudstrup}, \& {Sato}}]{hjorth2021}
{Hjorth}, M., {Albrecht}, S., {Hirano}, T., {et~al.} 2021, Proceedings of the
  National Academy of Science, 118, e2017418118,
  \dodoi{10.1073/pnas.2017418118}

\bibitem[{{Hogg} \& {Foreman-Mackey}(2018)}]{hogg2018}
{Hogg}, D.~W., \& {Foreman-Mackey}, D. 2018, \apjs, 236, 11,
  \dodoi{10.3847/1538-4365/aab76e}

\bibitem[{{Holcomb} {et~al.}(2022){Holcomb}, {Robertson}, {Hartigan},
  {Oelkers}, \& {Robinson}}]{holcomb2022}
{Holcomb}, R.~J., {Robertson}, P., {Hartigan}, P., {Oelkers}, R.~J., \&
  {Robinson}, C. 2022, \apj, 936, 138, \dodoi{10.3847/1538-4357/ac8990}

\bibitem[{{Huehnerhoff} {et~al.}(2016){Huehnerhoff}, {Ketzeback}, {Bradley},
  {Dembicky}, {Doughty}, {Hawley}, {Johnson}, {Klaene}, {Leon}, {McMillan},
  {Owen}, {Sayres}, {Sheen}, \& {Shugart}}]{huehnerhoff2016}
{Huehnerhoff}, J., {Ketzeback}, W., {Bradley}, A., {et~al.} 2016, in \procspie,
  Vol. 9908, , 99085H, \dodoi{10.1117/12.2234214}

\bibitem[{{Hunter}(2007)}]{hunter2007}
{Hunter}, J.~D. 2007, Computing in Science and Engineering, 9, 90,
  \dodoi{10.1109/MCSE.2007.55}

\bibitem[{{Ichimoto} \& {Kurokawa}(1984)}]{ichimoto1984}
{Ichimoto}, K., \& {Kurokawa}, H. 1984, \solphys, 93, 105,
  \dodoi{10.1007/BF00156656}

\bibitem[{{Jenkins} {et~al.}(2016){Jenkins}, {Twicken}, {McCauliff},
  {Campbell}, {Sanderfer}, {Lung}, {Mansouri-Samani}, {Girouard}, {Tenenbaum},
  {Klaus}, {Smith}, {Caldwell}, {Chacon}, {Henze}, {Heiges}, {Latham},
  {Morgan}, {Swade}, {Rinehart}, \& {Vanderspek}}]{jenkins2016}
{Jenkins}, J.~M., {Twicken}, J.~D., {McCauliff}, S., {et~al.} 2016, in
  \procspie, Vol. 9913, , 99133E, \dodoi{10.1117/12.2233418}

\bibitem[{{Kanodia} \& {Wright}(2018)}]{kanodia2018}
{Kanodia}, S., \& {Wright}, J. 2018, RNAAS, 2, 4,
  \dodoi{10.3847/2515-5172/aaa4b7}

\bibitem[{{Kipping}(2013)}]{kipping2013}
{Kipping}, D.~M. 2013, \mnras, 435, 2152, \dodoi{10.1093/mnras/stt1435}

\bibitem[{Kluyver {et~al.}(2016)Kluyver, Ragan-Kelley, P{\'e}rez, Granger,
  Bussonnier, Frederic, Kelley, Hamrick, Grout, Corlay, Ivanov, Avila, Abdalla,
  Willing, \& development team}]{jupyter2016}
Kluyver, T., Ragan-Kelley, B., P{\'e}rez, F., {et~al.} 2016, in Positioning and
  Power in Academic Publishing: Players, Agents and Agendas, ed. F.~Loizides \&
  B.~Scmidt (IOS Press), 87--90.
\newblock \url{https://eprints.soton.ac.uk/403913/}

\bibitem[{{Kreidberg}(2015)}]{kreidberg2015}
{Kreidberg}, L. 2015, \pasp, 127, 1161, \dodoi{10.1086/683602}

\bibitem[{{Lightkurve Collaboration} {et~al.}(2018){Lightkurve Collaboration},
  {Cardoso}, {Hedges}, {Gully-Santiago}, {Saunders}, {Cody}, {Barclay}, {Hall},
  {Sagear}, {Turtelboom}, {Zhang}, {Tzanidakis}, {Mighell}, {Coughlin}, {Bell},
  {Berta-Thompson}, {Williams}, {Dotson}, \& {Barentsen}}]{lightkurve}
{Lightkurve Collaboration}, {Cardoso}, J.~V.~d.~M., {Hedges}, C., {et~al.}
  2018, {Lightkurve: Kepler and TESS time series analysis in Python}, ASCL.
\newblock \doeprint{1812.013}

\bibitem[{{Masuda} \& {Winn}(2020)}]{masuda2020}
{Masuda}, K., \& {Winn}, J.~N. 2020, \aj, 159, 81,
  \dodoi{10.3847/1538-3881/ab65be}

\bibitem[{McCully {et~al.}(2018)McCully, Crawford, Kovacs, Tollerud, Betts,
  Bradley, Craig, Turner, Streicher, Sipocz, Robitaille, \&
  Deil}]{astroscrappy}
McCully, C., Crawford, S., Kovacs, G., {et~al.} 2018,
  \dodoi{10.5281/zenodo.1482019}

\bibitem[{McKinney(2010)}]{pandas2010}
McKinney, W. 2010, in Proceedings of the 9th Python in Science Conference, ed.
  S.~van~der Walt \& J.~Millman, 51 -- 56

\bibitem[{{McLaughlin}(1924)}]{mclaughlin1924}
{McLaughlin}, D.~B. 1924, \apj, 60, \dodoi{10.1086/142826}

\bibitem[{{Morris} {et~al.}(2018){Morris}, {Tollerud}, {Sip{\H o}cz}, {Deil},
  {Douglas}, {Berlanga Medina}, {Vyhmeister}, {Smith}, {Littlefair},
  {Price-Whelan}, {Gee}, \& {Jeschke}}]{morris2018}
{Morris}, B.~M., {Tollerud}, E., {Sip{\H o}cz}, B., {et~al.} 2018, \aj, 155,
  128, \dodoi{10.3847/1538-3881/aaa47e}

\bibitem[{{Naoz}(2016)}]{naoz2016}
{Naoz}, S. 2016, \araa, 54, 441, \dodoi{10.1146/annurev-astro-081915-023315}

\bibitem[{{Osborn} {et~al.}(2022){Osborn}, {Bonfanti}, {Gandolfi}, {Hedges},
  {Leleu}, {Fortier}, {Futyan}, {Gutermann}, {Maxted}, {Borsato}, {Collins},
  {Gomes da Silva}, {G{\'o}mez Maqueo Chew}, {Hooton}, {Lendl}, {Parviainen},
  {Salmon}, {Schanche}, {Serrano}, {Sousa}, {Tuson}, {Ulmer-Moll}, {Van
  Grootel}, {Wells}, {Wilson}, {Alibert}, {Alonso}, {Anglada}, {Asquier},
  {Barrado y Navascues}, {Baumjohann}, {Beck}, {Benz}, {Biondi}, {Bonfils},
  {Bouchy}, {Brandeker}, {Broeg}, {B{\'a}rczy}, {Barros}, {Cabrera}, {Charnoz},
  {Collier Cameron}, {Csizmadia}, {Davies}, {Deleuil}, {Delrez}, {Demory},
  {Ehrenreich}, {Erikson}, {Fossati}, {Fridlund}, {Gillon}, {G{\"o}mez-Munoz},
  {G{\"u}del}, {Heng}, {Hoyer}, {Isaak}, {Kiss}, {Laskar}, {Lecavelier des
  Etangs}, {Lovis}, {Magrin}, {Malavolta}, {McCormac}, {Nascimbeni},
  {Olofsson}, {Ottensamer}, {Pagano}, {Pall{\'e}}, {Peter}, {Piazza}, {Piotto},
  {Pollacco}, {Queloz}, {Ragazzoni}, {Rando}, {Rauer}, {Reimers}, {Ribas},
  {Demangeon}, {Smith}, {Sabin}, {Santos}, {Scandariato}, {Schroffenegger},
  {Schwarz}, {Shporer}, {Simon}, {Steller}, {Szab{\'o}}, {S{\'e}gransan},
  {Thomas}, {Udry}, {Walter}, \& {Walton}}]{osborn2022}
{Osborn}, H.~P., {Bonfanti}, A., {Gandolfi}, D., {et~al.} 2022, \aap, 664,
  A156, \dodoi{10.1051/0004-6361/202243065}

\bibitem[{Parviainen(2016)}]{pyde}
Parviainen, H. 2016, PyDE: v1.5, \dodoi{10.5281/zenodo.45602}

\bibitem[{{Pepper} {et~al.}(2007){Pepper}, {Pogge}, {DePoy}, {Marshall},
  {Stanek}, {Stutz}, {Poindexter}, {Siverd}, {O'Brien}, {Trueblood}, \&
  {Trueblood}}]{pepper2017}
{Pepper}, J., {Pogge}, R.~W., {DePoy}, D.~L., {et~al.} 2007, \pasp, 119, 923,
  \dodoi{10.1086/521836}

\bibitem[{{Petigura}(2015)}]{petigura2015}
{Petigura}, E. 2015, arXiv e-prints, arXiv:1510.03902.
\newblock \doarXiv{1510.03902}

\bibitem[{{Petrovich} {et~al.}(2020){Petrovich}, {Mu{\~n}oz}, {Kratter}, \&
  {Malhotra}}]{petrovich2020}
{Petrovich}, C., {Mu{\~n}oz}, D.~J., {Kratter}, K.~M., \& {Malhotra}, R. 2020,
  \apjl, 902, L5, \dodoi{10.3847/2041-8213/abb952}

\bibitem[{{Rasio} \& {Ford}(1996)}]{rasio1996}
{Rasio}, F.~A., \& {Ford}, E.~B. 1996, Science, 274, 954,
  \dodoi{10.1126/science.274.5289.954}

\bibitem[{{Ricker} {et~al.}(2015){Ricker}, {Winn}, {Vanderspek}, {Latham},
  {Bakos}, {Bean}, {Berta-Thompson}, {Brown}, {Buchhave}, {Butler}, {Butler},
  {Chaplin}, {Charbonneau}, {Christensen-Dalsgaard}, {Clampin}, {Deming},
  {Doty}, {De Lee}, {Dressing}, {Dunham}, {Endl}, {Fressin}, {Ge}, {Henning},
  {Holman}, {Howard}, {Ida}, {Jenkins}, {Jernigan}, {Johnson}, {Kaltenegger},
  {Kawai}, {Kjeldsen}, {Laughlin}, {Levine}, {Lin}, {Lissauer}, {MacQueen},
  {Marcy}, {McCullough}, {Morton}, {Narita}, {Paegert}, {Palle}, {Pepe},
  {Pepper}, {Quirrenbach}, {Rinehart}, {Sasselov}, {Sato}, {Seager},
  {Sozzetti}, {Stassun}, {Sullivan}, {Szentgyorgyi}, {Torres}, {Udry}, \&
  {Villasenor}}]{ricker2015}
{Ricker}, G.~R., {Winn}, J.~N., {Vanderspek}, R., {et~al.} 2015, JATIS, 1,
  014003, \dodoi{10.1117/1.JATIS.1.1.014003}

\bibitem[{{Rossiter}(1924)}]{rossiter1924}
{Rossiter}, R.~A. 1924, \apj, 60, \dodoi{10.1086/142825}

\bibitem[{{Rubenzahl} {et~al.}(2021){Rubenzahl}, {Dai}, {Howard}, {Chontos},
  {Giacalone}, {Lubin}, {Rosenthal}, {Isaacson}, {Batalha}, {Crossfield},
  {Dressing}, {Fulton}, {Huber}, {Kane}, {Petigura}, {Robertson}, {Roy},
  {Weiss}, {Beard}, {Hill}, {Mayo}, {Mocnik}, {Murphy}, \&
  {Scarsdale}}]{rubenzahl2021}
{Rubenzahl}, R.~A., {Dai}, F., {Howard}, A.~W., {et~al.} 2021, \aj, 161, 119,
  \dodoi{10.3847/1538-3881/abd177}

\bibitem[{Sanchis-Ojeda {et~al.}(2013)Sanchis-Ojeda, Winn, Marcy, Howard,
  Isaacson, Johnson, Torres, Albrecht, Campante, Chaplin, Davies, Lund, Carter,
  Dawson, Buchhave, Everett, Fischer, Geary, Gilliland, Horch, Howell, \&
  Latham}]{sanchisojeda2013}
Sanchis-Ojeda, R., Winn, J.~N., Marcy, G.~W., {et~al.} 2013, The Astrophysical
  Journal, 775, 54, \dodoi{10.1088/0004-637x/775/1/54}

\bibitem[{{Sasaki} \& {Suto}(2021)}]{sasaki2021}
{Sasaki}, S., \& {Suto}, Y. 2021, \pasj, 73, 1656, \dodoi{10.1093/pasj/psab102}

\bibitem[{{Schwab} {et~al.}(2016){Schwab}, {Rakich}, {Gong}, {Mahadevan},
  {Halverson}, {Roy}, {Terrien}, {Robertson}, {Hearty}, {Levi}, {Monson},
  {Wright}, {McElwain}, {Bender}, {Blake}, {St{\"u}rmer}, {Gurevich},
  {Chakraborty}, \& {Ramsey}}]{schwab2016}
{Schwab}, C., {Rakich}, A., {Gong}, Q., {et~al.} 2016, in \procspie, Vol. 9908,
  , 99087H, \dodoi{10.1117/12.2234411}

\bibitem[{{Southworth}(2011)}]{southworth2011}
{Southworth}, J. 2011, \mnras, 417, 2166,
  \dodoi{10.1111/j.1365-2966.2011.19399.x}

\bibitem[{{Speagle}(2020)}]{speagle2019}
{Speagle}, J.~S. 2020, \mnras, 493, 3132, \dodoi{10.1093/mnras/staa278}

\bibitem[{{Stefansson} {et~al.}(2017){Stefansson}, {Mahadevan}, {Hebb},
  {Wisniewski}, {Huehnerhoff}, {Morris}, {Halverson}, {Zhao}, {Wright},
  {O'rourke}, {Knutson}, {Hawley}, {Kanodia}, {Li}, {Hagen}, {Liu}, {Beatty},
  {Bender}, {Robertson}, {Dembicky}, {Gray}, {Ketzeback}, {McMillan}, \&
  {Rudyk}}]{stefansson2017}
{Stefansson}, G., {Mahadevan}, S., {Hebb}, L., {et~al.} 2017, \apj, 848, 9,
  \dodoi{10.3847/1538-4357/aa88aa}

\bibitem[{{Stefansson} {et~al.}(2018){Stefansson}, {Mahadevan}, {Wisniewski},
  {Li}, {Hebb}, {Morris}, {Halverson}, {Monson}, \&
  {Robertson}}]{stefansson2018b}
{Stefansson}, G., {Mahadevan}, S., {Wisniewski}, J., {et~al.} 2018, in
  \procspie, Vol. 10702, G, 1070250, \dodoi{10.1117/12.2312833}

\bibitem[{{Stefansson} {et~al.}(2020){Stefansson}, {Mahadevan}, {Maney},
  {Ninan}, {Robertson}, {Rajagopal}, {Haase}, {Allen}, {Ford}, {Winn},
  {Wolfgang}, {Dawson}, {Wisniewski}, {Bender}, {Ca{\~n}as}, {Cochran},
  {Diddams}, {Fredrick}, {Halverson}, {Hearty}, {Hebb}, {Kanodia}, {Levi},
  {Metcalf}, {Monson}, {Ramsey}, {Roy}, {Schwab}, {Terrien}, \&
  {Wright}}]{stefansson2020b}
{Stefansson}, G., {Mahadevan}, S., {Maney}, M., {et~al.} 2020, \aj, 160, 192,
  \dodoi{10.3847/1538-3881/abb13a}

\bibitem[{{Stefánsson} {et~al.}(2022){Stefánsson}, {Mahadevan}, {Petrovich},
  {Winn}, {Kanodia}, {Millholland}, {Maney}, {Ca{\~n}as}, {Wisniewski},
  {Robertson}, {Ninan}, {Ford}, {Bender}, {Blake}, {Cegla}, {Cochran},
  {Diddams}, {Dong}, {Endl}, {Fredrick}, {Halverson}, {Hearty}, {Hebb},
  {Hirano}, {Lin}, {Logsdon}, {Lubar}, {McElwain}, {Metcalf}, {Monson},
  {Rajagopal}, {Ramsey}, {Roy}, {Schwab}, {Schweiker}, {Terrien}, \&
  {Wright}}]{stefansson2022}
{Stefánsson}, G., {Mahadevan}, S., {Petrovich}, C., {et~al.} 2022, \apjl, 931,
  L15, \dodoi{10.3847/2041-8213/ac6e3c}

\bibitem[{van~der Walt {et~al.}(2011)van~der Walt, Colbert, \&
  Varoquaux}]{vanderwalt2011}
van~der Walt, S., Colbert, S.~C., \& Varoquaux, G. 2011, Computing in Science
  and Engineering, 13, 22, \dodoi{10.1109/MCSE.2011.37}

\bibitem[{{Winn} {et~al.}(2010){Winn}, {Johnson}, {Howard}, {Marcy},
  {Isaacson}, {Shporer}, {Bakos}, {Hartman}, \& {Albrecht}}]{winn2010}
{Winn}, J.~N., {Johnson}, J.~A., {Howard}, A.~W., {et~al.} 2010, \apjl, 723,
  L223, \dodoi{10.1088/2041-8205/723/2/L223}

\bibitem[{Wirth {et~al.}(2021)Wirth, Zhou, Quinn, Mann, Bouma, Latham, Teske,
  Wang, Shectman, Butler, \& Crane}]{wirth2021}
Wirth, C.~P., Zhou, G., Quinn, S.~N., {et~al.} 2021, \apjl, 917, L34,
  \dodoi{10.3847/2041-8213/ac13a9}

\bibitem[{{Yee} {et~al.}(2017){Yee}, {Petigura}, \& {von Braun}}]{yee2017}
{Yee}, S.~W., {Petigura}, E.~A., \& {von Braun}, K. 2017, \apj, 836, 77,
  \dodoi{10.3847/1538-4357/836/1/77}

\bibitem[{{Zechmeister} {et~al.}(2018){Zechmeister}, {Reiners}, {Amado},
  {Azzaro}, {Bauer}, {B{\'e}jar}, {Caballero}, {Guenther}, {Hagen}, {Jeffers},
  {Kaminski}, {K{\"u}rster}, {Launhardt}, {Montes}, {Morales}, {Quirrenbach},
  {Reffert}, {Ribas}, {Seifert}, {Tal-Or}, \& {Wolthoff}}]{zechmeister2018}
{Zechmeister}, M., {Reiners}, A., {Amado}, P.~J., {et~al.} 2018, \aap, 609,
  A12, \dodoi{10.1051/0004-6361/201731483}

\bibitem[{{Zhang} {et~al.}(2023){Zhang}, {Knutson}, {Dai}, {Wang}, {Ricker},
  {Schwarz}, {Mann}, \& {Collins}}]{zhang2022}
{Zhang}, M., {Knutson}, H.~A., {Dai}, F., {et~al.} 2023, \aj, 165, 62,
  \dodoi{10.3847/1538-3881/aca75b}

\end{thebibliography}
\bibliographystyle{aasjournal}

\end{document}